\documentclass[twocolumn,showpacs,preprintnumbers,amsmath,amssymb,
showkeys,floatfix]{revtex4}
\usepackage{graphicx}
\usepackage{dcolumn}
\usepackage{bm}
\usepackage{bigints}
\usepackage{natbib}

\begin{document}

\title{Stokes parameters spectral distortions due to the Sunyaev-Zel'dovich effect and an independent estimation of the CMB low multipoles}

\author{D.~I. Novikov}
\affiliation{Astro-Space Center of P.N. Lebedev Physical Institute, Profsoyusnaya 84/32, Moscow, Russia 117997}

\affiliation{Moscow Institute of Physics and Technology, Institutskiy Pereulok, d.9, Dolgoprudny, Moscow Region, Russia}

\author{S.~V. Pilipenko}
\affiliation{Astro-Space Center of P.N. Lebedev Physical Institute, Profsoyusnaya 84/32, Moscow, Russia 117997}

\author{M. De Petris}
\affiliation{Department of Physics, Sapienza University of Rome, Piazzale Aldo Moro 2, Rome, I-00185 Italy}

\author{G. Luzzi}
\affiliation{ASI Space Science Data Center, Via del Politecnico snc Rome, 00133 Italy}
\affiliation{Department of Physics, Sapienza, University of Rome, Piazzale Aldo Moro 2, Rome, I-00185 Italy}

\author{A.~O. Mihalchenko}
\affiliation{Moscow Institute of Physics and Technology, Institutskiy Pereulok, d.9, Dolgoprudny, Moscow Region, Russia}


\begin{abstract}
  
  We consider the Stokes parameters' frequency spectral distortions arising
  due to Compton scattering of the anisotropic cosmic microwave background
  (CMB) radiation, the Sunyaev-Zel’dovich effect (SZ), towards clusters of
  galaxies. We single out a
  very special type of such distortions and find simple analytical
  formulas for them. We show that this kind of distortion has a very
  distinctive spectral shape and can be separated from other kinds of
  contaminants.
  We demonstrate that this effect gives us an opportunity for an independent
  estimation of the low-multipole angular CMB anisotropies, such as the
  dipole, the quadrupole, and the octupole.
  We also show that, using distorted signals from nearby and distant clusters,
  one can distinguish between the Sachs-Wolfe and the
  integrated Sachs-Wolfe effects. The detection of such distortions
  can be feasible with high-angular resolution and high-sensitivity space
  missions, such as the upcoming Millimetron Space Observatory experiment.
\end{abstract}


\maketitle

\section{Introduction}

The interaction of cosmic microwave background (CMB) radiation with hot
plasma in clusters of galaxies causes the Comptonization of relic photons.
This process is commonly known as the thermal Sunyaev-Zel'dovich effect (tSZ)
\citep{1969Ap&SS...4..301Z,1970Ap&SS...7...20S}. As a result
of this effect, a deviation in the frequency spectrum of 
radiation from the original CMB blackbody shape occurs. Another reason for the
change in the CMB spectrum is the so-called kinematic Sunyaev-Zel'dovich
effect (kSZ), which arises as a result of the peculiar cluster movements
with respect to the CMB rest frame as well as the intracluster medium (ICM) motions, such as coherent gas rotations \citep{2018MNRAS.479.4028B} or
substructure merging processes \citep{2017A&A...598A.115A}. Nowadays, it is clear that
SZ observations with high spatial
and spectral capabilities allow us to deeply exploit clusters'
ICM distribution even in the low-mass and high-redshift regimes, so complementing x-ray-band data. Correct
knowledge of the thermodynamics of the cluster atmosphere is required to use them
as cosmological tools and to possibly disentangle high-order-effect signals
\citep{2013MNRAS.430.3054C, 2019MNRAS.490..784R} such as the
ones we are interested in. The SZ effect is a
very effective tool for the investigation of the internal structure of galaxies 
and galaxy clusters \cite{2011A&A...535A.108C,2011A&A...527L...1C,2000A&A...360..417E,2015MNRAS.452.1328M}. It also has cosmological applications
for the investigation of dark matter and dark energy
\citep{2004A&A...422L..23C,2002PhRvL..88w1301W}, reionization
\citep{2006PhRvD..73j3001C,2016A&A...595A..21C}, and the independent estimation
of the Hubble constant \citep{1978ApJ...226L.103S,1979MNRAS.187..847B,1977ApJ...217....6C} and CMB temperature; see Refs.
\citep{2015JCAP...09..011L, 2012ApJ...757..144D}.
In Refs. \citep{2002ARA&A..40..643C,2019SSRv..215...17M}, the detailed overview
of the SZ effect and its use for cosmology was provided. These articles
discussed the observational results, theoretical and observational challenges,
various techniques, new opportunities and possible new directions
in this field.

In the past thirty years, a large number of theoretical
investigations connected with various corrections to the Kompaneets equation \citep{1957JETP...4.730K} (which describes the photon spectrum evolution in electron plasma) and the SZ effect have been performed. The generalization of the Kompaneets equation, relativistic corrections to the thermal SZ effect and multiple scattering
on SZ clusters were considered in Refs. \citep{2012MNRAS.425.1129C,2014MNRAS.437...67C,2014MNRAS.438.1324C,1998ApJ...499....1C,1998ApJ...502....7I,Stebbins:1997qr,Rephaeli:2002zs,2000astro.ph..5390I,1994PhRvD..49..648H}. A
detailed study of relativistic correction was made in Refs. 
\citep{1998ApJ...508...17N,1998ApJ...508....1S,1999ApJ...510..930C} for
kinematic SZ. In Refs.  \citep{2009PhRvD..79h3005N,2013MNRAS.434..710N,2014MNRAS.441.3018N}, a detailed analytical investigation
of the Boltzmann equations was made for three Lorentz frames and expressions for the photon redistribution functions were derived. An effective way to
separate kinematic and scattering terms was obtained in Ref. \citep{2012MNRAS.426..510C}. The influence of the motion of the Solar System on the SZ signal as
an additional correction was discussed in Ref.  \citep{2005A&A...434..811C}.

In addition to the change in the intensity spectrum, the SZ effect also causes
linear polarization, mainly due to the presence of the quadrupole component of
CMB anisotropy in the galaxy cluster location \citep{2004MNRAS.347..729L,1999MNRAS.310..765S}. This
polarization is produced as a result of  ``cold"  Thomson scattering in the non-relativistic regime. This effect was considered in Ref. \citep{1997PhRvD..56.4511K} for
clusters at high redshifts to measure the CMB quadrupole, thereby getting
around cosmic variance.
In the relativistic regime other multipoles of CMB anisotropy can also contribute to the
polarization of scattered radiation  \citep{2000MNRAS.312..159C,2016JCAP...07..031S}. In addition to this, kSZ can also produce polarization
\citep{2000ApJ...533..588I}.
The influence of the CMB anisotropy on the spectral distortions for intensity and polarization in tSZ/kSZ was considered in Ref. \citep{2016PhRvD..94b3513Y}. In that work, it was shown how low multipoles form the spectrum of Stokes parameters in radiation scattered by moving clusters. 

Therefore, Compton scattering of an anisotropic blackbody radiation on hot relativistic plasma creates some particular spectral features in
both: radiation intensity $I$ and linear polarization parameters $Q$ and $U$.
The leading term in the intensity distortion is the classical thermal SZ effect
caused by the isotropic (monopole $\ell=0$) fraction of incident radiation. In addition to this, CMB anisotropy multipoles with $\ell=1,2,3$ influence the intensity of
scattered radiation, causing the distortion of a very special shape \citep{2018PhRvD..98l3513E}. Moreover, in the relativistic regime, polarization parameters caused
by scattering will also have a very characteristic frequency of spectral features,
but unlike intensity, this effect occurs due to the presence of the CMB quadrupole
($\ell=2$) and octupole ($\ell=3$).

The effects described above are in fact anisotropic corrections to the classical thermal SZ effect and follow directly from the anisotropic Kompaneets equation first derived in Ref. \citep{1969JPhys..30..301B} and generalized for polarization in Refs.  \citep{1974ApJ...191..183P,1981MNRAS.195..115S} in linear approximation in $kT_e/m_ec^2$. 

The amplitudes of $I,Q,U$ distortions depend on the powers of CMB $C_1, C_2, C_3$ multipoles and their orientation with respect to the axis connecting the scattering point and the observer. 
For the observer located at a nearby SZ cluster the CMB anisotropy map is approximately the same as we directly observe in the sky, including the quadrupole and octupole. According to Ref.  \citep{2018PhRvD..98l3513E}, at redshifts
$z\sim 0.05$ the variations of multipole
amplitudes are about $10\%$. Therefore, the measurements of intensity and polarization from such clusters allow us to estimate $\ell=1,2,3$ CMB multipoles
amplitudes and their orientations.
Since WMAP and Planck results show insufficient power in low-CMB anisotropy
multipoles \citep{2003MNRAS.346L..26E,2003PhRvD..68l3523T,2004PhRvL..93v1301S,2014A&A...571A..15P} and quadrupole-octupole alignment \citep{2004PhRvD..70d3515C,2006MNRAS.367...79C,2008IJMPD..17..179N} (which is difficult to explain statistically), it is important to have an independent source
of information for the estimation of $\ell=1,2,3$ multipoles. Besides, we will be
able to get the information about the intrinsic dipole in our location.

The detection of such a signal from distant clusters can provide us with information about the large-angular-scale CMB anisotropy radiation coming directly from the surface of the last scattering  [pure Sachs-Wolfe (SW) effect] without the influence of Rees-Sciama (RS) \citep{1967ApJ...147...73S} or integrated Sachs-Wolfe (ISW) effect.

In our paper, we derive simple analytical formulas for spectral distortions of
Stokes parameters due to anisotropic tSZ effect and select the special component of these distortions which is responsible for the nonblackbody fraction of radiation
(we show that all other components change the temperature, leaving the spectrum in the blackbody shape). The ratio of the
amplitude of this very distinctive component to the amplitude of classical tSZ
is not sensitive to the temperature or the density of the ICM and depends only on the linear combination of CMB local multipoles with
$\ell=1,2,3$. We do not take into account cluster movement because our
expected signal can be easily separated from the components associated with the kSZ effect. We show how to use distorted
signals of this kind coming from
nearby and distant clusters to estimate $\ell=1,2,3$ CMB multipoles and to
distinguish between the SW and ISW effects.

It is worth mentioning that the signal we consider is very low and should be
separated from other types of spectral distortions. As for intensity distortion, 
we have to take into account that the CMB isotropic monopole part is larger than
anisotropic multipoles by a factor of $\sim 10^5$. This means that we should
take into account relativistic corrections of higher orders to this effect.
At the same time, the polarised signal is not affected by the CMB monopole. Therefore, 
$Q,U$ observational data will be ``cleaner" than the data for intensity.

The future Millimetron mission  \citep{2009ExA....23..221W,2014PhyU...57.1199K,2012SPIE.8442E..4CS} will have unprecedented sensitivity in the single-dish mode to measure the polarized signal in the range from 100 GHz to 2 THz down to sub-arcminute resolution. One of the main tasks of this mission will be the measurement of $\mu$ and $y$ spectral distortions of CMB radiation, which arise due to the energy injections into the plasma in the prerecombination epoch. These distortions can also be formed
as a result of the interaction of CMB with matter during the process of large-scale structure formation  
\citep{1969Ap&SS...4..301,1970Ap&SS...7...20S,1974AZh....51.1162I,1991A&A...246...49B,1993PhRvD..48..485H,2012MNRAS.419.1294C,2015MNRAS.454.4182C,2019BAAS...51c.184C,2019arXiv190901593C,2015MNRAS.451.4460D,2014arXiv1405.6938C}. As an additional task, the possibility of detecting the spectral distortions from SZ clusters caused by CMB anisotropy can be considered. We discuss the feasibility of this task by
Millimetron in Sec. IV (Conclusions).

The  outline of this paper is as follows: In Sec. II, we
review the equation for the polarized radiative transfer in plasma
in the relativistic regime. Using this equation, we derive analytical formulas for
$I,Q,U$ distortions for scattered radiation and find expressions for
observed linear combinations of multipole amplitudes.
In Sec. III, we show how to estimate low CMB multipoles and
how to distinguish between SW and ISW effects
combining the signals from nearby and distant clusters.
Finally, in Sec. IV, we make our brief conclusions.

\section{Stokes parameters spectral distortions due to anisotropic thermal
  SZ effect}

 In this section we review the radiative polarization transfer equations in hot
 Comptonizing electron plasma, which was first derived by Stark in Ref. \citep{1981MNRAS.195..115S}, and we apply
 this approach to the consideration of CMB scattering on Sunyaev-Zel'dovich
 clusters. Combining results found in Refs. \citep{1969JPhys..30..301B} and \citep{1981MNRAS.195..115S} and assuming that the incident radiation is unpolarized,
 we find the simple analytical
 formulas for Stokes parameters spectral distortions that arise due to the
 scattering of anisotropic radiation on hot electrons.\\
 
 \vspace{0.2cm}
 
 We use the following notations:\\
 
 \vspace{0.2cm}
 
$I(\nu,\bold{\Omega}),Q(\nu,\bold{\Omega}),U(\nu,\bold{\Omega})$:
Stokes parameters, where $I$ is the spectral radiance, $Q$ and $U$
describe the linear polarization, $\nu$ is the frequency and
$\bold{\Omega}$ is the radiation propagation direction.\\

\vspace{0.2cm}

$n=\frac{c^2I}{2h\nu^3}$: Photon concentration in phase
space, where $c$ is the speed of light and $h$ is the Planck constant.
We also use the notations $q=\frac{c^2Q}{2h\nu^3}$ and
$u=\frac{c^2U}{2h\nu^3}$.\\

\vspace{0.2cm}

${\bf J}=\begin{pmatrix}n(\nu,\bold{\Omega})\\q(\nu,\bold{\Omega})\\
u(\nu,\bold{\Omega})\end{pmatrix}$ is the vector that describes
Stokes parameters.

\vspace{0.2cm}

$T_e, T_r$: Temperatures of electrons and radiation.\\

\vspace{0.2cm}

$m_e$: Electron mass at rest.\\

\vspace{0.2cm}

$\Theta_e=\frac{kT_e}{m_ec^2}$, $\varepsilon=\frac{h\nu}{m_ec^2}$,
$x=\frac{h\nu}{kT_r}$, where $k$ is the Boltzmann constant.

\begin{figure}[htb]
  \includegraphics[width=7cm]{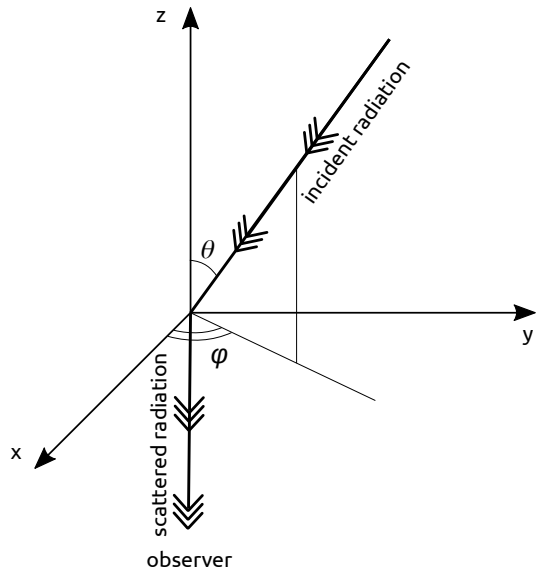}
  \caption{Schematic representation of radiation scattering on a SZ cluster.
    The scattering
    point (SZ cluster) is at the center of the coordinate system. This system
    is chosen so that the \textit{z} axis is directed from the observer to the cluster.
    The scattered radiation is propagating to the observer in the direction
    opposite to the \textit{z} axis. Therefore, $\Theta$ is the angle of scattering, and
    $\varphi$ denotes the scattering plane orientation.}
\end{figure}
\begin{figure}[htb]
  \includegraphics[width=7cm]{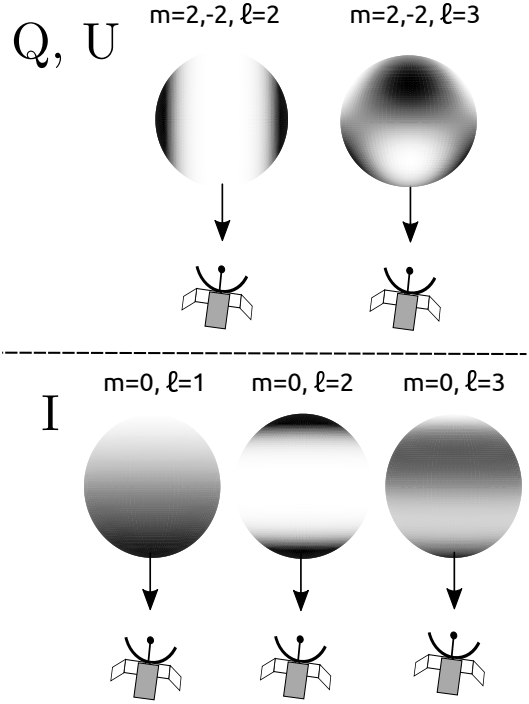}
  \caption{Schematic representation of CMB multipoles $Y_{\ell}^m$ in the
    surroundings of the cluster location that contribute to the spectral distortions of type 2.
    Here the \textit{z} axis points from the observer to the cluster in the corresponding
    spherical coordinate system.
    \textit{Upper panel}: $Q,U$ are affected by $\ell=2,3$ and $m=\pm2$.
    \textit{Lower panel}: Intensity $I$ is distorted by axisymmetric multipoles
    $\ell=1,2,3$, $m=0$.}
\end{figure}

The equation for the radiative transfer in plasma for anisotropic radiation
was first derived in Ref.  \citep{1969JPhys..30..301B} and generalized for the polarized radiation in Ref. \citep{1981MNRAS.195..115S}. It can be
used for relatively cold photons and electrons: $kT_e\ll m_ec^2$,
$h\nu\ll m_ec^2$. This approximation works pretty well for SZ clusters
with $T_e\sim 10$ keV and CMB temperature
$T_r\sim 3$ K. This approach implies the correction
to the Thomson scattering up to the first order in $\varepsilon$, $\Theta_e$.
In our investigation, we assume an optically thin limit, so that photons propagating through the cluster can scatter
only once, and the radiation before scattering is unpolarized and anisotropic.
Thus, if we denote the Stokes parameters by ``prime" values $n^{{}_{'}},q^{{}_{'}},u^{{}_{'}}$ before scattering,
then $q^{{}_{'}}=u^{{}_{'}}=0$. In order to separate the isotropic and anisotropic parts of radiation, it is convenient to use
$\bar{n}(\nu)=\frac{1}{4\pi}\int n^{{}_{'}}(\nu,\bold{\Omega})d\bold{\Omega}$ and 
$\Delta=n^{{}_{'}}(\nu,\bold{\Omega})-\bar{n}(\nu)$.

Using our notations and omitting some obvious algebra in the case of single
scattering, we can rewrite the polarization transfer equation in hot
Comptonizing electron plasma found in Ref.  \citep{1981MNRAS.195..115S} as follows:
 
\begin{equation}
  \begin{array}{l}
    \vspace{0.2cm}
    \delta{\bf J}/\delta\tau=\frac{3}{16\pi}\bigint\limits_{\bold{\Omega}}
    \Big\{\left.\Delta{\bf R}+
    \Theta_e\Delta\left({\bf S}-2\frac{T_r}{T_e}{\bf R}\right)\right.+\\
    \vspace{0.5cm}
    \left.\Theta_eG_a(\Delta)(1-\mu){\bf R}\right.\Big\}d\bold{\Omega}
    +\Theta_eG_i(\bar{n}){\bf T}+O(\Theta^2),\\
    
    \vspace{0.5cm}
    
    {\bf R}=\begin{pmatrix}1+\mu^2\\(1-\mu^2)\cos2\varphi\\
    (1-\mu^2)\sin2\varphi\end{pmatrix},\hspace{0.5cm}
    {\bf T}=\begin{pmatrix}1\\0\\0\end{pmatrix}\\
    
    \vspace{0.5cm}
    
    {\bf S}=2\begin{pmatrix}(1-\mu)^2(1+2\mu)-4\mu\\
    (1-\mu^2)(1+2\mu)\cos2\varphi\\
    (1-\mu^2)(1+2\mu)\sin2\varphi\end{pmatrix},\hspace{0.2cm}d\bold{\Omega}=
    d\mu d\varphi\\
    
    \vspace{0.5cm}
    
    G_a(\Delta)=\frac{1}{x^2}\frac{\partial}{\partial x}
  \left[ x^4\left(\frac{T_r}{T_e}\Delta+
    \frac{\partial\Delta}{\partial x}\right)\right]
  +2\frac{T_r}{T_e}n^{{}_{'}}
  \frac{\partial}{\partial x}\left[x^2\Delta\right],\\
  
  \vspace{0.5cm}
  
   G_i(\bar{n})=\frac{1}{x^2}\frac{\partial}{\partial x}\left[x^4\left(\frac{T_r}{T_e}
  (\bar{n}+\bar{n}^2)+\frac{\partial \bar{n}}{\partial x}\right)\right].
  
 \end{array}
\end{equation}
Here $\delta\tau$ is the optical depth, and $\mu=\cos\Theta$ is the cosine of the scattering angle. The azimuthal angle
$\varphi$ shows the orientation of the scattering plane (Fig 1).
Functions $G_a$ and $G_i$ represent the anisotropic and isotropic parts
of radiation, respectively. Since the radiation temperature is much less
than the temperature of plasma in galaxy clusters one can neglect all the
terms proportional to $T_r/T_e$ in Eq. (1):
\begin{equation}
  \begin{array}{l}
    \vspace{0.2cm}
    \delta{\bf J}/\delta\tau=\\

    \vspace{0.5cm}
    
    \frac{3}{16\pi}\bigint\limits_{\bold{\Omega}}
    \Big\{\Delta\left({\bf R}+
    \Theta_e{\bf S}\right)+
    G(\Delta)\Theta_e(1-\mu){\bf R}
    \Big\}d\bold{\Omega}\\

    \vspace{0.5cm}
    
    +\Theta_eG(\bar{n}){\bf T}+O(\Theta^2),
    \hspace{0.5cm} 
    G(f)=\frac{1}{x^2}\frac{\partial}{\partial x}
    \left[x^4\frac{\partial f}{\partial x}\right],
 \end{array}
\end{equation}
with $f$ equal to $\Delta$ or $\bar{n}$.
The CMB radiation frequency spectrum can be
considered as a blackbody spectrum with high accuracy \citep{2009ApJ...707..916F} . Therefore, the shape
of the spectrum can be completely described by the radiation temperature $T_r$.
Since the relic radiation is anisotropic, its temperature varies with direction. Thus, for the radiation incident on a cluster we can write the following
formulas:

\begin{equation}
  \begin{array}{l}
    
    \vspace{0.5cm}
    
    n^{{}_{'}}(x,\bold{\Omega})=B(x)-x\frac{dB}{dx}
    \hspace{0.1cm}\frac{\Delta_T(\bold{\Omega})}{T_r},\hspace{0.5cm}
    B(x)=\frac{1}{e^x-1}, \\

     \vspace{0.5cm}

    \Delta_T(\bold{\Omega})=T(\bold{\Omega})-T_r,\\
    
    \vspace{0.5cm}
    
    \bar{n}(x)=B(x),\hspace{1cm}    
    \Delta=-x\frac{dB}{dx}
    \hspace{0.1cm}\frac{\Delta_T(\bold{\Omega})}{T_r}.

    \end{array}
\end{equation}
Substituting these formulas into Eq. (2), we get very simple 
analytical results for Stokes parameters spectral distortions after a
single scattering:
\begin{equation}
  \begin{array}{l}
    \vspace{0.2cm}
    \delta{\bf J}/\delta\tau=\\

     \vspace{0.5cm}
    
    g_1(x)\left[\frac{3}{16\pi}\int\limits_{\bold{\Omega}}\frac{\Delta_T(\bold{\Omega})}
      {T_r}\left({\bf R}+\Theta_e{\bf S}\right) d\bold{\Omega}\right]+
    \hspace{0.4cm}\Big\}1\\
    
    \vspace{0.5cm}
    
   +\Theta_eg_2(x)\left[\frac{3}{16\pi}\int\limits_{\bold{\Omega}}\frac{\Delta_T(\bold{\Omega})}
    {T_r}(1-\mu){\bf R}d\bold{\Omega}\right]+\hspace{0.2cm}\Big\}2\\

    \vspace{0.5cm}

    +\Theta_eg_3(x){\bf T},\hspace{3cm}\Big\}3\hspace{0.2cm}(tSZ)\\

     \vspace{0.5cm}
    
     g_1(x)=-x\frac{dB}{dx},\hspace{0.5cm}g_2(x)=\frac{1}{x^2}\frac{d}{dx}
     \left[x^4\frac{d}{dx}\left(-x\frac{dB}{dx}\right)\right],\\
     
      \vspace{0.5cm}
     
     g_3(x)=\frac{1}{x^2}\frac{d}{dx}
    \left[x^4\frac{dB}{dx}\right]
 \end{array}
\end{equation}

The expression for $\delta J/\delta\tau$ consists of three terms.
They represent spectral distortions of three different types
$g_1$, $g_2$, and $g_3$.
Below, we describe the physical meaning of these three terms and analyze
the possibility to use them for the investigation of CMB properties.

1. {\bf The first term} represents Thompson scattering with some correction
proportional to $\Theta_e$. This term arises due to mixing flows
with different temperatures in the process of scattering and
represents distortion
of the form $g_1=-x\frac{dB}{dx}$. It is rather difficult to use
distortion of this kind in real observations for the
following reasons:\\
a. The first Stokes parameter $I$ (or $n$) after scattering preserves
its spectral shape in the form of a blackbody but with a bit different
temperature. Indeed, in the linear approximation the term
$-x\frac{dB}{dx}\frac{\Delta_T}{T_r}$ is
in fact the difference between two blackbody spectra with temperatures
$T_r$ and $T_r+\Delta T$. Thus the amplitude of such deviation depends
on the value of the mean temperature $T_r$.\\
b. The CMB is slightly polarized and its linear
polarization parameters have a spectral shape in the form of $g_1$. Therefore, Stokes
parameters $q$ and $u$ of the radiation propagating through the
cluster to the observer without scattering have
the same spectral shape as the Stokes parameters for the scattered fraction
of radiation. Thus, scattered and unscattered fractions can be confused with each other if we use this type of polarization spectral distortion.

2. {\bf The second term} represents distortion of the form
$g_2(x)=\frac{1}{x^2}\frac{d}{dx}\left
[x^4\frac{d}{dx}\left(-x\frac{dB}{dx}\right)\right]$. This effect occurs
due to Comptonization of anisotropic incident radiation on hot electrons.
Such distortion has a very characteristic and distinctive nonblackbody
shape for
all three Stokes parameters. Using distortions of this type gives us a
unique opportunity to separate the scattered fraction of radiation
from the unscattered one. In our previous paper  \citep{2018PhRvD..98l3513E}, we analyzed this
distortion only for the first Stokes parameter $n$, calling this effect
an ``anisotropic thermal SZ effect."

3. {\bf The third term} is the classical thermal SZ effect:
$g_3(x)=\frac{1}{x^2}\frac{d}{dx}\left[x^4\frac{dB}{dx}\right]$. This term
represents the distortion for the isotropic part of incident radiation. It 
affects the $n$ parameter only, without modifying linear polarization:
\begin{equation}
\delta\bold{J_3}=\begin{pmatrix}\delta n_{{}_3}\\\delta q_{{}_3}\\\delta u_{{}_3}\end{pmatrix}=
g_3(x)\begin{pmatrix}1\\0\\0\end{pmatrix}\Theta_e\delta\tau,
\end{equation}
where the index 3 means that we take into account distortion of the form $g_3$.

Let us consider how low multipoles of CMB anisotropy contribute
to the Stokes parameters spectral distortions of the form $g_2(x)$. This
is the distortion which is of particular interest to us and the only type
of distortion we will analyze below. We will denote this part of the distortion
by the index 2.

The temperature fluctuations of radiation incident to the SZ cluster can be described
in terms of spherical harmonics $Y_{\ell}^m(\bold{\Omega})$:
\begin{equation} 
  \frac{\Delta_T(\bold{\Omega})}{T_r}=\sum\limits_{\ell=1}^{\infty}
  \sum\limits_{m=-\ell}^{\ell}\tilde{a}_{\ell,m}Y_{\ell}^m(\bold{\Omega}),
\end{equation}
where $\tilde{a}_{\ell,m}$ are the coefficients for $\Delta_T/T_r$ decomposition
for a given location of the galaxy cluster ($\tilde{a}_{\ell, m}$ are not the same
as $a_{\ell,m}$ for the conventional Galactic coordinate system). We choose a local spherical coordinate system at the scattering point in such a way that the south pole corresponds
to the direction from the SZ cluster to the observer (see Fig. 1). Substituting Eq. (6) into the second term of Eq. (4), we get

\begin{equation}
  \begin{array}{l}

    \vspace{0.3cm}
    
   \frac{\delta n_{{}_2}}{\delta\tau}=-\frac{3}{4\sqrt{\pi}}\left(\frac{4}{5\sqrt{3}}\tilde{a}_{1,0}-
    \frac{1}{3\sqrt{5}}\tilde{a}_{2,0}+
    \frac{1}{5\sqrt{7}}\tilde{a}_{3,0}\right)\Theta_eg_2(x),\\

    \vspace{0.3cm}
    
    \frac{\delta q_{{}_2}}{\delta\tau}=\frac{3}{4\sqrt{5\pi}}\left(\tilde{a}_{2,2}-
    \frac{1}{\sqrt{7}}\tilde{a}_{3,2}\right)\Theta_eg_2(x),\\

     \vspace{0.3cm}
    
    \frac{\delta u_{{}_2}}{\delta\tau}=\frac{3}{4\sqrt{5\pi}}\left(\tilde{a}_{2,-2}-
    \frac{1}{\sqrt{7}}\tilde{a}_{3,-2}\right)\Theta_eg_2(x).
    
  \end{array}
\end{equation}
Therefore, only multipoles with $\ell=1,2,3$ contribute to this kind of distortion. In Fig. 2, you can see the schematic demonstration of the local CMB anisotropy
multipoles that contribute to the distortion of the observed Stokes
parameters. The first parameter $n$ is distorted by axisymmetric local
multipoles $Y_{\ell}^0$, $\ell=1,2,3$, while the linear polarization parameters
$q$ and $u$ are distorted due to the presence of multipoles with $m=\pm2$:
$Y_2^{\pm2}$ and $Y_3^{\pm2}$. It is worth noticing, that the orientation of the
distorted part of linear polarization is different from the orientation of
polarization caused by pure ``cold" Thomson scattering. It happens because
Thomson scattering produces linear polarization due to the presence of
harmonics $\ell=2$, $m=\pm2$ in the anisotropy, while the distorted part of the
polarization is produced by harmonics with $\ell=2,3$,\hspace{0.2cm}$m=\pm2$.
We can write the following formulas:
{\large
\begin{equation*}
  \begin{array}{l}

    \vspace{0.2cm}
    
    \tan(2\psi_2)=\frac{u_2}{q_2}=
    \frac{\tilde{a}_{2,-2}-\frac{1}{\sqrt{7}}\tilde{a}_{3,-2}}
         {\tilde{a}_{2,2}-\frac{1}{\sqrt{7}}\tilde{a}_{3,2}},\\

         \vspace{0.2cm}

    \tan(2\psi_{cold})=\frac{u_{cold}}{q_{cold}}=\frac{\tilde{a}_{2,-2}}
         {\tilde{a}_{2,2}},
 \end{array}
\end{equation*}
\par}where $\psi_2$ and $\psi_{cold}$ are the distorted polarization orientation
and the orientation of the polarization caused by Thomson scattering respectively.
We can always rotate the coordinate system in such a way that
$\hat{u}_{cold}=0$ and $\psi_{cold}=0$. Therefore, we can estimate the angle between
the orientations of the Thomson polarization and the distorted part of the linear polarization in terms of coefficients $\hat{a}_{ell,m}$ in the rotated (``hat") coordinate
system:
{\large
\begin{equation*}
  \begin{array}{l}
    \psi_2-\psi_{cold}=-\frac{1}{2}
    \arctan\left(\frac{\hat{a}_{3,-2}}
         {\sqrt{7}\hat{a}_{2,2}-\hat{a}_{3,2}}\right).\\
 \end{array}
\end{equation*}
\par}

In order to estimate the CMB anisotropy multipoles at the SZ cluster
location we should divide the amplitude of the observed distorted signal
proportional to $g_2$ by the amplitude of the classical thermal SZ effect,
which is proportional to $g_3$ [Eq. 5].
In this case, we can get rid of the Comptonization parameter
$\Theta_e\delta_{\tau}$ and find the following coefficients
$\beta$:

\begin{equation}
  \begin{array}{l}

    \vspace{0.3cm}
    
   \beta_n=-\frac{3}{4\sqrt{\pi}}\left(\frac{4}{5\sqrt{3}}\tilde{a}_{1,0}-
    \frac{1}{3\sqrt{5}}\tilde{a}_{2,0}+
    \frac{1}{5\sqrt{7}}\tilde{a}_{3,0}\right),\\

    \vspace{0.3cm}
    
    \beta_q=\frac{3}{4\sqrt{5\pi}}\left(\tilde{a}_{2,2}-
    \frac{1}{\sqrt{7}}\tilde{a}_{3,2}\right),\\

     \vspace{0.3cm}
    
    \beta_u=\frac{3}{4\sqrt{5\pi}}\left(\tilde{a}_{2,-2}-
    \frac{1}{\sqrt{7}}\tilde{a}_{3,-2}\right).
    
  \end{array}
\end{equation}
Therefore, by observing distorted radiation coming from a SZ cluster
one can find three coefficients $\beta_n$, $\beta_q$ and $\beta_u$
that are the linear combinations of the local harmonics amplitudes
with $\ell=1,2,3$ and $-\ell\le m\le \ell$.

We should mention that the intensity spectral distortions are affected
by the isotropic part of CMB radiation, which is 5 orders of magnitude
larger than the anisotropic one. At the same time, the linear
polarization induced by the SZ effect arises due to the anisotropic fraction
only. This means that in the process of component separation for intensity,
we should take into account relativistic corrections to tSZ of higher
orders \citep{1998ApJ...499....1C}. In addition to this, relativistic
corrections due to bulk motion of hot clusters should be also taken into
account \citep{1998ApJ...508....1S}. Thermal corrections are proportional
to $\tau\Theta_e^k$, $k>1$ (see Fig 3). The observational data for
polarization will be free of such corrections, and therefore will be ``cleaner" than the intensity data.

It is worth noticing, that the anisotropy in the cluster medium together
with the multiple scattering can induce an additional local radiation
anisotropy. When CMB scatters on a cluster's medium, the radiation frequency
spectrum becomes distorted mainly due to the tSZ effect. As a result of
this effect, radiation incident to the point of the last scattering
before propagating to the observer receives an additional anisotropy.
This induced anisotropy $\Delta_{in}$ is the anisotropy
of the tSZ amplitude (not the temperature anisotropy of a blackbody
radiation). Such an anisotropy arises due to different
medium temperatures $\Theta_e(\bold{\Omega})$ and optical depths
$\delta_{\tau}(\bold{\Omega})$ in different directions
$\bold{\Omega}$ from the point of the last scattering:
\begin{equation*}
  \Delta_{in}(x,\bold{\Omega})=\frac{1}{x^2}\frac{d}{dx}
  \left[x^4\frac{dB}{dx}\right]\Theta_e(\bold{\Omega})\delta_{\tau}
  (\bold{\Omega})
\end{equation*}
[compare with $\Delta$ in Eq. (3)]. Therefore, this kind of anisotropy will
create an additional spectral distortion, but the shape of such a
distortion will be different from what we are considering.

\begin{figure}[!htbp]
  \includegraphics[width=1\columnwidth]{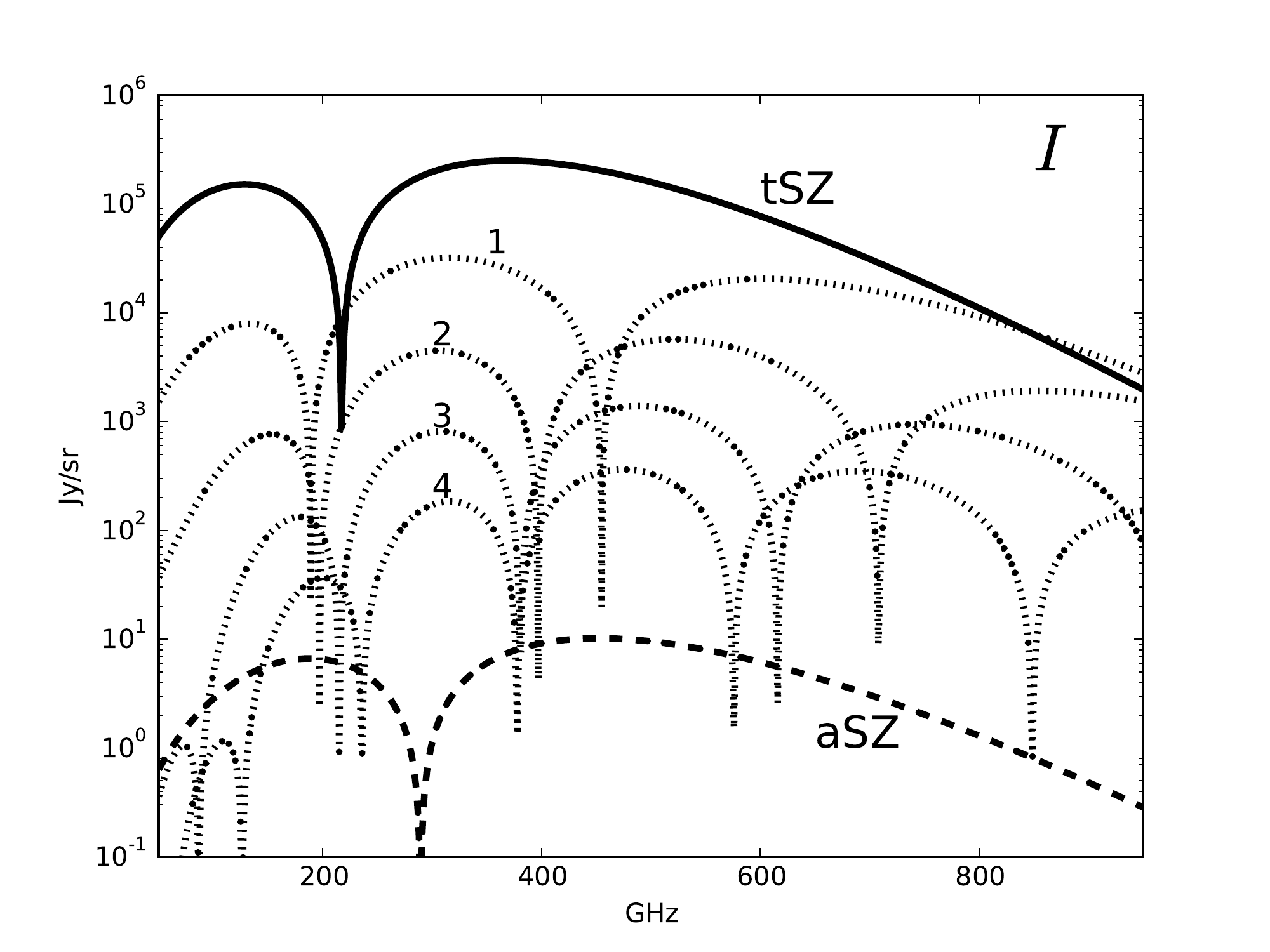}
  \includegraphics[width=1\columnwidth]{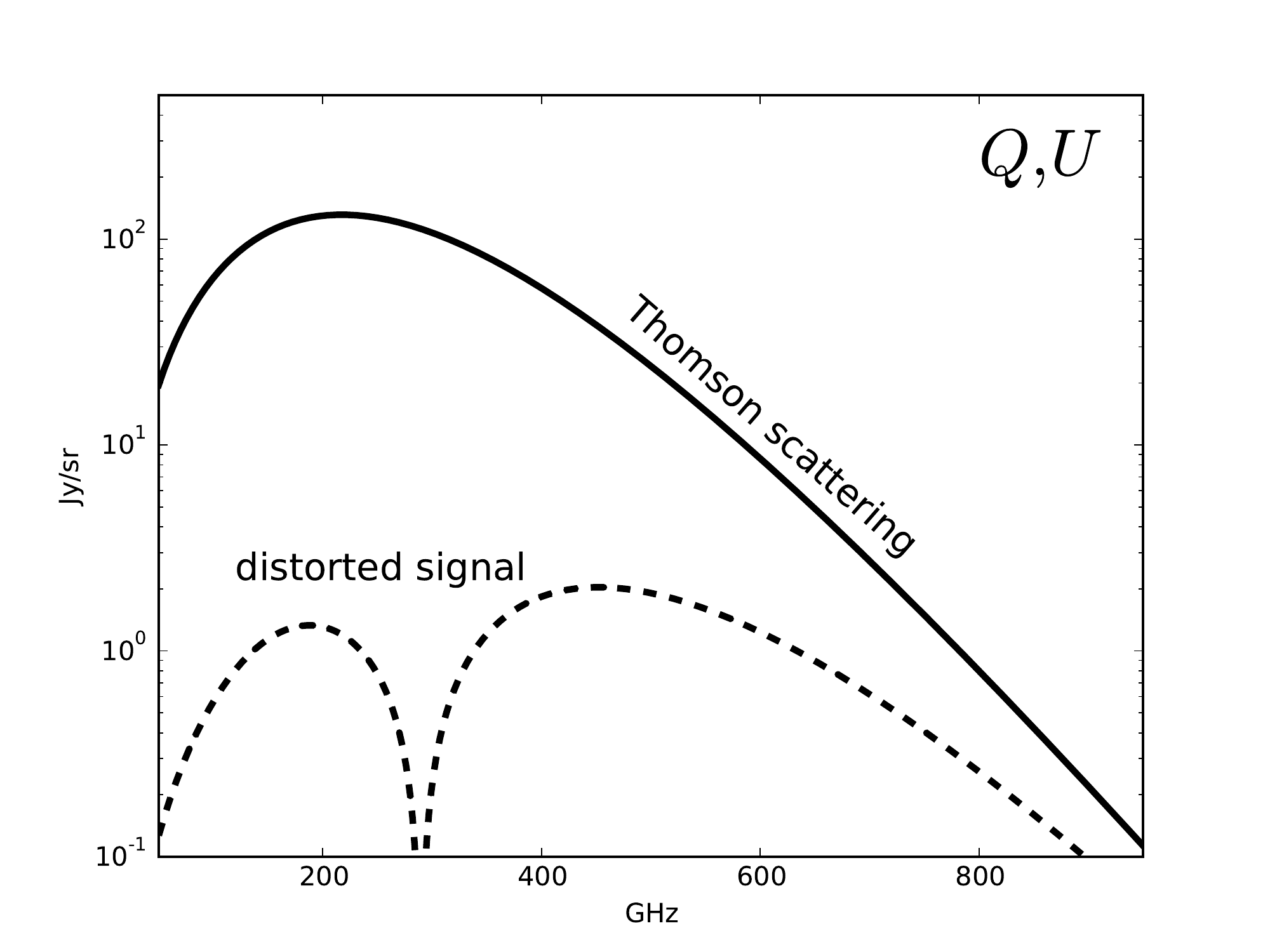}
  \caption{A distorted signal of type 2 for the intensity and polarization
    from
    a cluster with optical depth $\tau=0.01$ and temperature $T_e=7$ keV.
    \textit{Upper panel}: The solid line shows the classical tSZ effect, dotted lines
    show relativistic corrections to tSZ up to the fourth order, and the 
    dashed line corresponds to the type 2 intensity distortion
    [``anisotropic SZ effect" (aSZ)]. Note that the aSZ crosses zero at
    a different frequency respecting the tSZ zero-crossing point. The
    amplitude and the sign of the aSZ effect depend on $\Theta_e\delta\tau$
    and the linear combination of $a_{\ell m}$ [see Eq 7].
    \textit{Lower panel}: The solid line shows the spectrum for
    the linear polarization caused by ``cold" Thomson scattering, and the dashed line
    is the spectrum of the type 2 distortion of the polarized fraction.}
\end{figure}

In the next section, we show how to use Eq. (8) for the independent
estimation of low CMB multipoles using nearby clusters. We will
also demonstrate how to distinguish between the SW and ISW
effects combining the signals from nearby and distant clusters.

\section{The independent estimation of low-CMB anisotropy multipoles and
  possibile separation of SW and ISW effects}

    In this section, we describe the method to reconstruct the CMB dipole, quadrupole,
    and octupole at our location and show how to separate contributions from the
    SW and ISW effects on CMB anisotropy by observing
    the distorted signals from SZ galaxy clusters.
    
As it follows from Eq. (8), the spectral distortion intensity depends on the linear combination of components of $\ell=1,2,3$ spherical harmonics, while the polarized signal depends on components with $\ell=2,3$. For both polarized and unpolarized signals, the components which contribute to the signal are projected on the axis pointing from the observer to the cluster. If the coefficients of spherical harmonics, $a_{\ell, m}$, were the same over the Universe, then by observing different clusters, it would be possible to measure different projections of these spherical harmonics, and thus reconstruct the full set of coefficients for $\ell=1,2,3$.

In reality, $a_{\ell, m}$ are not the same over the space, because low-$\ell$ harmonics are significantly affected by the ISW effect. According to Ref. \cite{1996PhRvL..76..575C}, about 40\% of the quadrupole amplitude and 25\% of the octupole amplitude are generated by the ISW effect. In our previous paper \citep{2018PhRvD..98l3513E} we analyzed the spatial correlations of $a_{\ell, m}$ and concluded that at distances less than 250~$ h^{-1}$\,Mpc from the Earth, the coefficients may differ no more than 10\% from the $a_{\ell, m}$ which we would observe in the absence of foregrounds. This will allow us to reconstruct local $a_{\ell, m}$ by observing $\sim$170 nearby clusters at $z<0.085$ (which corresponds to 250~$h^{-1}$\,Mpc).

In Figs. 4 and 5, we show the maps of the expected intensity and polarization
distortions of type 2 for the scattered radiation coming from nearby SZ clusters located at particular positions in the sky. These maps correspond to the values of $a_{\ell, m}$ for $\ell=2,3$ measured by \textit{Planck} \citep{2018arXiv180706208P}. In the case of intensity, we did not take into account the intrinsic CMB dipole, since it is completely overshadowed by
our motion with respect to the CMB rest frame.
In Fig. 4, we show the celestial map in Galactic coordinates, where the value of the intensity distortion is indicated by the shade of gray, and lines correspond to the orientation and amplitude of polarization, determined by $\beta_q$ and $\beta_u$. In Fig. 5, we show the
amplitude of the distorted polarized signal, i.e., $\sqrt{\beta_q^2+\beta_u^2}$. A lighter color corresponds to stronger polarization. 

\begin{figure}[tp]
	\includegraphics[width=1\columnwidth]{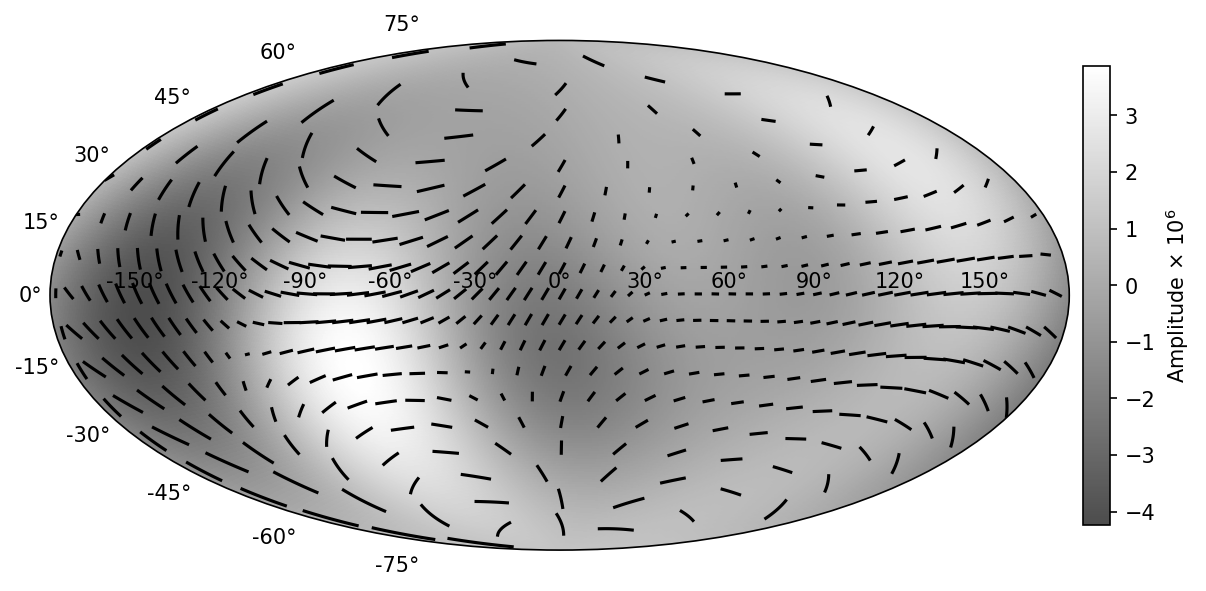}
	\caption{Map of the unpolarized signal amplitude, shown with gray color. The polarized signal determined by $\beta_q$ and $\beta_u$ is shown by short lines. The map is based on values of $a_{\ell, m}$ given by the Planck Collaboration \citep{2018arXiv180706208P}.}
\end{figure}

\begin{figure}[tp]
	\includegraphics[width=1\columnwidth]{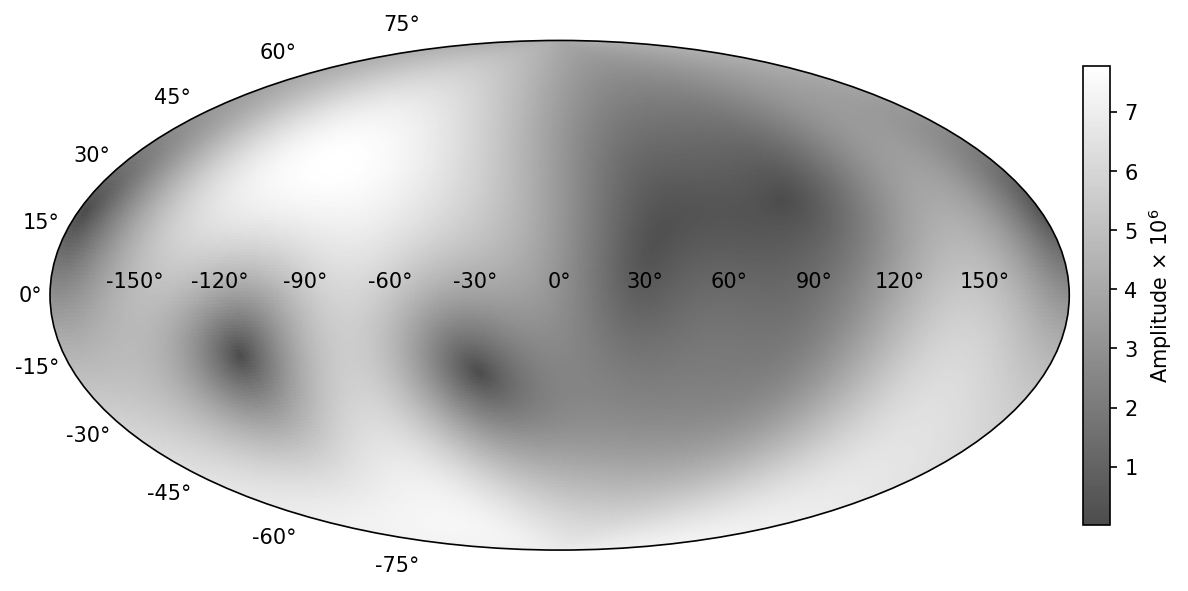}
	\caption{Map of polarization amplitude, $\sqrt{\beta_q^2+\beta_u^2}$, based on values of $a_{\ell, m}$ given by the Planck Collaboration \citep{2018arXiv180706208P}. Lighter colors correspond to stronger polarization.}
\end{figure}

If we look at distances $\ge$\,$1000\,h^{-1}$\,Mpc, the ISW effect is weak, and low-$\ell$ anisotropy of CMB is created by the SW effect on the last scattering surface. The coefficients $a_{\ell, m}$ weakly depend on the redshift, if we consider clusters at distances 1000$\,h^{-1}$\,Mpc $<R<$ 2000$\,h^{-1}$\,Mpc. Reconstruction of $a_{\ell, m}$ at these distances will allow us to measure the low-$\ell$ CMB anisotropy at the last scattering surface, i.e., without the ISW effect.

Now we will show how the reconstruction of $a_{\ell, m}$ from measurements of $\beta_n$, $\beta_q$ and $\beta_u$ for a set of clusters can be done, and how measurements of polarization can improve the precision of reconstruction. We imply that $a_{\ell, m}$ are the same for the window of distances we consider. The signal in Eq. (8) depends on the values of $\tilde{a}_{1,0}$, $\tilde{a}_{2,0}$, $\tilde{a}_{3,0}$, $\tilde{a}_{2,\pm 2}$, $\tilde{a}_{3,\pm 2}$, where the tilde denotes that the coefficients are given in a rotated coordinate system. The rotation needed to obtain $\tilde{a}_{\ell, m}$ from $a_{\ell, m}$ is defined as follows: For a cluster with Galactic sky coordinates $l, b$ we first rotate the system around \textit{z} axis by $-l$, and then rotate about the new \textit{y} axis by $90-b$ degrees. For any cluster location, we can express the three signals ($\beta_n$, $\beta_q$, $\beta_u$) as a linear combination of all 15 $a_{\ell, m}$'s with $\ell=1,2,3$. The coefficients in this linear combination depend on the celestial coordinates. As a result, for $N$ different celestial locations we have a set of $3N$ linear equations. The values of $a_{\ell, m}$ can be determined by fitting a linear model to the measured signals.

We can estimate the relative precision of $a_{\ell, m}$ determination for known SZ clusters from the PLANCKSZ2 catalog  \citep{2016A&A...594A..27P} by using the maximum likelihood estimator. We start with 15 clusters at $R>1000$ $h^{-1}$ Mpc, as this is the minimum number of clusters needed to find $a_{\ell, m}$ in the case where we measure only the unpolarized signal, $\beta_n$. We set the error of $a_{\ell, m}$ determination to 1.0 for this sample of 15 clusters and the case when only $\beta_n$ is used. We grow the sample by adding more clusters at larger distances, and we consider either using only the unpolarized signal, or the full signal. The results are shown in Fig. 6. As one can see from this figure, the precision of octupole and quadrupole determination is improved by a factor of 3, when the polarization measurements are used.

\begin{figure}[t]
	\includegraphics[width=1\columnwidth]{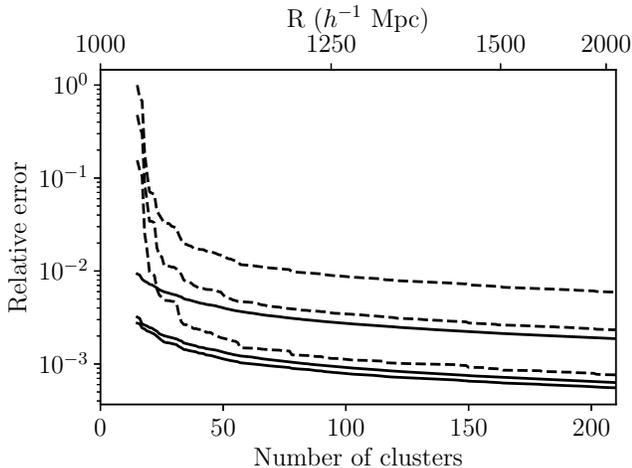}
	\caption{Relative error of determination of amplitudes of dipole, quadrupole and octupole (from bottom to top) with respect to an error for a minimal sample of 15 clusters, as a function of the cumulative number of clusters in a sample growing from 1000$\,h^{-1}$ to 2000$\,h^{-1}$\,Mpc. Solid lines correspond to the usage of the full (polarized + unpolarized) signal; dashed lines correspond to only the unpolarized signal ($\beta_n$).}
\end{figure}

Observing two samples of clusters, one at $R<250\,h^{-1}$\,Mpc and another one at $R>1000$ $h^{-1}$\,Mpc, will allow us to distinguish between SW and ISW effects. Using the first sample, we will determine local $a_{\ell, m}$ created by both SW and ISW effects, but free from the contamination by zodiacal light and our Galaxy's dust. By measuring the signal from distant clusters in the second sample, we will reconstruct the $a_{\ell, m}$ created only by the SW effect. The relative precision for $a_{\ell, m}$ reconstruction will remain the same as in our previous paper \citep{2018PhRvD..98l3513E} if we only use the intensity distortions. We would remind the reader that the reconstruction of coefficients with $m=\ell$ is more
complicated. This is mainly because most of the power of such harmonics is close to the Galactic plane, where the number of clusters observed by \textit{Planck} is limited.

\section{Conclusions}

In our paper, we considered polarized thermal SZ effect and
derived an analytical result for a very distinctive component
of spectral distortions in Stokes parameters which arise
due to the presence of dipole, quadrupole, and octupole CMB
angular anisotropy. We showed that this type of distortion
can be disentangled from other components and can be used for
the independent estimation of $\ell=1,2,3$ CMB multipole
amplitudes and their orientation. We demonstrated that by
observing distorted radiation from nearby clusters it is possible to estimate independently CMB anisotropy coefficients
$a_{\ell, m}$, $\ell=1,2,3$, $-\ell\le m\le\ell$ in our location.
We also proposed a method for the separation of ISW and SW effects
by combining observations of distorted signals from distant and nearby clusters.

In this work, we did not discuss foreground component separation or the impact of kSZ, relativistic corrections, and CMB photon multiple
scattering.
The peculiar spectral shape of the type of distortions considered here allows us
to disentangle them from other contaminants.
The  anisotropies in the cluster medium together with the double scattering
can induce additional local anisotropy which should also be taken into account.
The large number of clusters in the same  redshift range and same sky direction
will provide us with good statistics, which can help to solve this problem.
Besides this, Millimetron, with its unprecedented
combination of high angular resolution and high sensitivity in a wide spectral
range will allow us to deeply observe individual clusters, providing accurate
maps of the ICM pressure distribution. We stress that under the assumption of a single scattering approximation, the ratio of the anisotropic SZ effect and the thermal one, aSZ/tSZ, is independent of the gas temperature and optical depth.

The signal we considered is strong enough for its detection with the
Millimetron mission to be feasible in principle.
Nevertheless, with today's
technologies, it implies a very long integration time. We should mention also
that accurate
knowledge of the instrumental polarization purity 
(cross polarization and/or instrumental induced polarization) is 
mandatory to avoid possible leakages between observed Stokes parameters, 
which can create irremovable systematic errors. It can affect our capability
to detect such a tiny spectral distortion in linear polarization.

\section{Acknowledgments}
We wish to thank the referee for very useful discussion.
The work is supported by the Project No. 01-2018 of LPI new scientific
groups and the Foundation for the Advancement of Theoretical Physics and Mathematics “BASIS”, Grant No. 19-1-1-46-1. M.D.P. acknowledges support from Sapienza Università di Roma thanks to Progetti di Ricerca Medi 2019, No. RM11916B7540DD8D.

\def\apj{Astrophys.~J}
\def\apjl{Astrophys.~J.,~Lett}
\def\apjs{Astrophys.~J.,~Supplement}
\def\an{Astron.~Nachr}
\def\aap{Astron.~Astrophys}
\def\mnras{Mon.~Not.~R.~Astron.~Soc}
\def\pasp{Publ.~Astron.~Soc.~Pac}
\def\aaps{Astron.~and Astrophys.,~Suppl.~Ser}
\def\apss{Astrophys.~Space.~Sci}
\def\ibvs{Inf.~Bull.~Variable~Stars}
\def\japa{J.~Astrophys.~Astron}
\def\na{New~Astron}
\def\aspproc{Proc.~ASP~conf.~ser.}
\def\aspcs{ASP~Conf.~Ser}
\def\aj{Astron.~J}
\def\actaa{Acta Astron}
\def\araa{Ann.~Rev.~Astron.~Astrophys}
\def\caosp{Contrib.~Astron.~Obs.~Skalnat{\'e}~Pleso}
\def\pasj{Publ.~Astron.~Soc.~Jpn}
\def\memsai{Mem.~Soc.~Astron.~Ital}
\def\astl{Astron.~Letters}
\def\aipproc{Proc.~AIP~conf.~ser.}
\def\physrep{Physics Reports}
\def\jcap{Journal of Cosmology and Astroparticle Physics}
\def\baas{Bulletin of the AAS}
\def\ssr{Space~Sci.~Rev.}
\def\azh{Astronomicheskii Zhurnal}

\bibliography{pol_resub2.bib}

\begin{thebibliography}{75}
\expandafter\ifx\csname natexlab\endcsname\relax\def\natexlab#1{#1}\fi
\expandafter\ifx\csname bibnamefont\endcsname\relax
  \def\bibnamefont#1{#1}\fi
\expandafter\ifx\csname bibfnamefont\endcsname\relax
  \def\bibfnamefont#1{#1}\fi
\expandafter\ifx\csname citenamefont\endcsname\relax
  \def\citenamefont#1{#1}\fi
\expandafter\ifx\csname url\endcsname\relax
  \def\url#1{\texttt{#1}}\fi
\expandafter\ifx\csname urlprefix\endcsname\relax\def\urlprefix{URL }\fi
\providecommand{\bibinfo}[2]{#2}
\providecommand{\eprint}[2][]{\url{#2}}

\bibitem[{\citenamefont{{Zeldovich} and
  {Sunyaev}}(1969{\natexlab{a}})}]{1969Ap&SS...4..301Z}
\bibinfo{author}{\bibfnamefont{Y.~B.} \bibnamefont{{Zeldovich}}}
  \bibnamefont{and} \bibinfo{author}{\bibfnamefont{R.~A.}
  \bibnamefont{{Sunyaev}}}, \bibinfo{journal}{\apss}
  \textbf{\bibinfo{volume}{4}}, \bibinfo{pages}{301}
  (\bibinfo{year}{1969}{\natexlab{a}}).

\bibitem[{\citenamefont{{Sunyaev} and {Zeldovich}}(1970)}]{1970Ap&SS...7...20S}
\bibinfo{author}{\bibfnamefont{R.~A.} \bibnamefont{{Sunyaev}}}
  \bibnamefont{and} \bibinfo{author}{\bibfnamefont{Y.~B.}
  \bibnamefont{{Zeldovich}}}, \bibinfo{journal}{\apss}
  \textbf{\bibinfo{volume}{7}}, \bibinfo{pages}{20} (\bibinfo{year}{1970}).

\bibitem[{\citenamefont{{Baldi} et~al.}(2018)\citenamefont{{Baldi}, {De
  Petris}, {Sembolini}, {Yepes}, {Cui}, and {Lamagna}}}]{2018MNRAS.479.4028B}
\bibinfo{author}{\bibfnamefont{A.~S.} \bibnamefont{{Baldi}}},
  \bibinfo{author}{\bibfnamefont{M.}~\bibnamefont{{De Petris}}},
  \bibinfo{author}{\bibfnamefont{F.}~\bibnamefont{{Sembolini}}},
  \bibinfo{author}{\bibfnamefont{G.}~\bibnamefont{{Yepes}}},
  \bibinfo{author}{\bibfnamefont{W.}~\bibnamefont{{Cui}}}, \bibnamefont{and}
  \bibinfo{author}{\bibfnamefont{L.}~\bibnamefont{{Lamagna}}},
  \bibinfo{journal}{\mnras} \textbf{\bibinfo{volume}{479}},
  \bibinfo{pages}{4028} (\bibinfo{year}{2018}), \eprint{1805.07142}.

\bibitem[{\citenamefont{{Adam} et~al.}(2017)\citenamefont{{Adam}, {Bartalucci},
  {Pratt}, {Ade}, {Andr{\'e}}, {Arnaud}, {Beelen}, {Beno{\^\i}t}, {Bideaud},
  {Billot} et~al.}}]{2017A&A...598A.115A}
\bibinfo{author}{\bibfnamefont{R.}~\bibnamefont{{Adam}}},
  \bibinfo{author}{\bibfnamefont{I.}~\bibnamefont{{Bartalucci}}},
  \bibinfo{author}{\bibfnamefont{G.~W.} \bibnamefont{{Pratt}}},
  \bibinfo{author}{\bibfnamefont{P.}~\bibnamefont{{Ade}}},
  \bibinfo{author}{\bibfnamefont{P.}~\bibnamefont{{Andr{\'e}}}},
  \bibinfo{author}{\bibfnamefont{M.}~\bibnamefont{{Arnaud}}},
  \bibinfo{author}{\bibfnamefont{A.}~\bibnamefont{{Beelen}}},
  \bibinfo{author}{\bibfnamefont{A.}~\bibnamefont{{Beno{\^\i}t}}},
  \bibinfo{author}{\bibfnamefont{A.}~\bibnamefont{{Bideaud}}},
  \bibinfo{author}{\bibfnamefont{N.}~\bibnamefont{{Billot}}},
  \bibnamefont{et~al.}, \bibinfo{journal}{\aap} \textbf{\bibinfo{volume}{598}},
  \bibinfo{eid}{A115} (\bibinfo{year}{2017}), \eprint{1606.07721}.

\bibitem[{\citenamefont{{Chluba} et~al.}(2013)\citenamefont{{Chluba},
  {Switzer}, {Nelson}, and {Nagai}}}]{2013MNRAS.430.3054C}
\bibinfo{author}{\bibfnamefont{J.}~\bibnamefont{{Chluba}}},
  \bibinfo{author}{\bibfnamefont{E.}~\bibnamefont{{Switzer}}},
  \bibinfo{author}{\bibfnamefont{K.}~\bibnamefont{{Nelson}}}, \bibnamefont{and}
  \bibinfo{author}{\bibfnamefont{D.}~\bibnamefont{{Nagai}}},
  \bibinfo{journal}{\mnras} \textbf{\bibinfo{volume}{430}},
  \bibinfo{pages}{3054} (\bibinfo{year}{2013}), \eprint{1211.3206}.

\bibitem[{\citenamefont{{Ruppin} et~al.}(2019)\citenamefont{{Ruppin}, {Mayet},
  {Mac{\'\i}as-P{\'e}rez}, and {Perotto}}}]{2019MNRAS.490..784R}
\bibinfo{author}{\bibfnamefont{F.}~\bibnamefont{{Ruppin}}},
  \bibinfo{author}{\bibfnamefont{F.}~\bibnamefont{{Mayet}}},
  \bibinfo{author}{\bibfnamefont{J.~F.} \bibnamefont{{Mac{\'\i}as-P{\'e}rez}}},
  \bibnamefont{and}
  \bibinfo{author}{\bibfnamefont{L.}~\bibnamefont{{Perotto}}},
  \bibinfo{journal}{\mnras} \textbf{\bibinfo{volume}{490}},
  \bibinfo{pages}{784} (\bibinfo{year}{2019}), \eprint{1905.05129}.

\bibitem[{\citenamefont{{Colafrancesco} and
  {Marchegiani}}(2011)}]{2011A&A...535A.108C}
\bibinfo{author}{\bibfnamefont{S.}~\bibnamefont{{Colafrancesco}}}
  \bibnamefont{and}
  \bibinfo{author}{\bibfnamefont{P.}~\bibnamefont{{Marchegiani}}},
  \bibinfo{journal}{\aap} \textbf{\bibinfo{volume}{535}}, \bibinfo{eid}{A108}
  (\bibinfo{year}{2011}), \eprint{1108.4602}.

\bibitem[{\citenamefont{{Colafrancesco}
  et~al.}(2011)\citenamefont{{Colafrancesco}, {Marchegiani}, and
  {Buonanno}}}]{2011A&A...527L...1C}
\bibinfo{author}{\bibfnamefont{S.}~\bibnamefont{{Colafrancesco}}},
  \bibinfo{author}{\bibfnamefont{P.}~\bibnamefont{{Marchegiani}}},
  \bibnamefont{and}
  \bibinfo{author}{\bibfnamefont{R.}~\bibnamefont{{Buonanno}}},
  \bibinfo{journal}{\aap} \textbf{\bibinfo{volume}{527}}, \bibinfo{eid}{L1}
  (\bibinfo{year}{2011}).

\bibitem[{\citenamefont{{En{\ss}lin} and {Kaiser}}(2000)}]{2000A&A...360..417E}
\bibinfo{author}{\bibfnamefont{T.~A.} \bibnamefont{{En{\ss}lin}}}
  \bibnamefont{and} \bibinfo{author}{\bibfnamefont{C.~R.}
  \bibnamefont{{Kaiser}}}, \bibinfo{journal}{\aap}
  \textbf{\bibinfo{volume}{360}}, \bibinfo{pages}{417} (\bibinfo{year}{2000}),
  \eprint{astro-ph/0001429}.

\bibitem[{\citenamefont{{Marchegiani} and
  {Colafrancesco}}(2015)}]{2015MNRAS.452.1328M}
\bibinfo{author}{\bibfnamefont{P.}~\bibnamefont{{Marchegiani}}}
  \bibnamefont{and}
  \bibinfo{author}{\bibfnamefont{S.}~\bibnamefont{{Colafrancesco}}},
  \bibinfo{journal}{\mnras} \textbf{\bibinfo{volume}{452}},
  \bibinfo{pages}{1328} (\bibinfo{year}{2015}), \eprint{1506.05651}.

\bibitem[{\citenamefont{{Colafrancesco}}(2004)}]{2004A&A...422L..23C}
\bibinfo{author}{\bibfnamefont{S.}~\bibnamefont{{Colafrancesco}}},
  \bibinfo{journal}{\aap} \textbf{\bibinfo{volume}{422}}, \bibinfo{pages}{L23}
  (\bibinfo{year}{2004}), \eprint{astro-ph/0405456}.

\bibitem[{\citenamefont{{Weller} et~al.}(2002)\citenamefont{{Weller}, {Battye},
  and {Kneissl}}}]{2002PhRvL..88w1301W}
\bibinfo{author}{\bibfnamefont{J.}~\bibnamefont{{Weller}}},
  \bibinfo{author}{\bibfnamefont{R.~A.} \bibnamefont{{Battye}}},
  \bibnamefont{and}
  \bibinfo{author}{\bibfnamefont{R.}~\bibnamefont{{Kneissl}}},
  \bibinfo{journal}{\prl} \textbf{\bibinfo{volume}{88}}, \bibinfo{eid}{231301}
  (\bibinfo{year}{2002}), \eprint{astro-ph/0110353}.

\bibitem[{\citenamefont{{Cooray}}(2006)}]{2006PhRvD..73j3001C}
\bibinfo{author}{\bibfnamefont{A.}~\bibnamefont{{Cooray}}},
  \bibinfo{journal}{\prd} \textbf{\bibinfo{volume}{73}}, \bibinfo{eid}{103001}
  (\bibinfo{year}{2006}), \eprint{astro-ph/0511240}.

\bibitem[{\citenamefont{{Colafrancesco}
  et~al.}(2016)\citenamefont{{Colafrancesco}, {Marchegiani}, and
  {Emritte}}}]{2016A&A...595A..21C}
\bibinfo{author}{\bibfnamefont{S.}~\bibnamefont{{Colafrancesco}}},
  \bibinfo{author}{\bibfnamefont{P.}~\bibnamefont{{Marchegiani}}},
  \bibnamefont{and} \bibinfo{author}{\bibfnamefont{M.~S.}
  \bibnamefont{{Emritte}}}, \bibinfo{journal}{\aap}
  \textbf{\bibinfo{volume}{595}}, \bibinfo{eid}{A21} (\bibinfo{year}{2016}),
  \eprint{1607.07723}.

\bibitem[{\citenamefont{{Silk} and {White}}(1978)}]{1978ApJ...226L.103S}
\bibinfo{author}{\bibfnamefont{J.}~\bibnamefont{{Silk}}} \bibnamefont{and}
  \bibinfo{author}{\bibfnamefont{S.~D.~M.} \bibnamefont{{White}}},
  \bibinfo{journal}{\apjl} \textbf{\bibinfo{volume}{226}},
  \bibinfo{pages}{L103} (\bibinfo{year}{1978}).

\bibitem[{\citenamefont{{Birkinshaw}}(1979)}]{1979MNRAS.187..847B}
\bibinfo{author}{\bibfnamefont{M.}~\bibnamefont{{Birkinshaw}}},
  \bibinfo{journal}{\mnras} \textbf{\bibinfo{volume}{187}},
  \bibinfo{pages}{847} (\bibinfo{year}{1979}).

\bibitem[{\citenamefont{{Cavaliere} et~al.}(1977)\citenamefont{{Cavaliere},
  {Danese}, and {de Zotti}}}]{1977ApJ...217....6C}
\bibinfo{author}{\bibfnamefont{A.}~\bibnamefont{{Cavaliere}}},
  \bibinfo{author}{\bibfnamefont{L.}~\bibnamefont{{Danese}}}, \bibnamefont{and}
  \bibinfo{author}{\bibfnamefont{G.}~\bibnamefont{{de Zotti}}},
  \bibinfo{journal}{\apj} \textbf{\bibinfo{volume}{217}}, \bibinfo{pages}{6}
  (\bibinfo{year}{1977}).

\bibitem[{\citenamefont{{Luzzi} et~al.}(2015)\citenamefont{{Luzzi},
  {G{\'e}nova-Santos}, {Martins}, {De Petris}, and
  {Lamagna}}}]{2015JCAP...09..011L}
\bibinfo{author}{\bibfnamefont{G.}~\bibnamefont{{Luzzi}}},
  \bibinfo{author}{\bibfnamefont{R.~T.} \bibnamefont{{G{\'e}nova-Santos}}},
  \bibinfo{author}{\bibfnamefont{C.~J.~A.~P.} \bibnamefont{{Martins}}},
  \bibinfo{author}{\bibfnamefont{M.}~\bibnamefont{{De Petris}}},
  \bibnamefont{and}
  \bibinfo{author}{\bibfnamefont{L.}~\bibnamefont{{Lamagna}}},
  \bibinfo{journal}{\jcap} \textbf{\bibinfo{volume}{2015}}, \bibinfo{eid}{011}
  (\bibinfo{year}{2015}), \eprint{1502.07858}.

\bibitem[{\citenamefont{{de Martino} et~al.}(2012)\citenamefont{{de Martino},
  {Atrio-Barandela}, {da Silva}, {Ebeling}, {Kashlinsky}, {Kocevski}, and
  {Martins}}}]{2012ApJ...757..144D}
\bibinfo{author}{\bibfnamefont{I.}~\bibnamefont{{de Martino}}},
  \bibinfo{author}{\bibfnamefont{F.}~\bibnamefont{{Atrio-Barandela}}},
  \bibinfo{author}{\bibfnamefont{A.}~\bibnamefont{{da Silva}}},
  \bibinfo{author}{\bibfnamefont{H.}~\bibnamefont{{Ebeling}}},
  \bibinfo{author}{\bibfnamefont{A.}~\bibnamefont{{Kashlinsky}}},
  \bibinfo{author}{\bibfnamefont{D.}~\bibnamefont{{Kocevski}}},
  \bibnamefont{and} \bibinfo{author}{\bibfnamefont{C.~J.~A.~P.}
  \bibnamefont{{Martins}}}, \bibinfo{journal}{\apj}
  \textbf{\bibinfo{volume}{757}}, \bibinfo{eid}{144} (\bibinfo{year}{2012}),
  \eprint{1203.1825}.

\bibitem[{\citenamefont{{Carlstrom} et~al.}(2002)\citenamefont{{Carlstrom},
  {Holder}, and {Reese}}}]{2002ARA&A..40..643C}
\bibinfo{author}{\bibfnamefont{J.~E.} \bibnamefont{{Carlstrom}}},
  \bibinfo{author}{\bibfnamefont{G.~P.} \bibnamefont{{Holder}}},
  \bibnamefont{and} \bibinfo{author}{\bibfnamefont{E.~D.}
  \bibnamefont{{Reese}}}, \bibinfo{journal}{\araa}
  \textbf{\bibinfo{volume}{40}}, \bibinfo{pages}{643} (\bibinfo{year}{2002}),
  \eprint{astro-ph/0208192}.

\bibitem[{\citenamefont{{Mroczkowski} et~al.}(2019)\citenamefont{{Mroczkowski},
  {Nagai}, {Basu}, {Chluba}, {Sayers}, {Adam}, {Churazov}, {Crites}, {Di
  Mascolo}, {Eckert} et~al.}}]{2019SSRv..215...17M}
\bibinfo{author}{\bibfnamefont{T.}~\bibnamefont{{Mroczkowski}}},
  \bibinfo{author}{\bibfnamefont{D.}~\bibnamefont{{Nagai}}},
  \bibinfo{author}{\bibfnamefont{K.}~\bibnamefont{{Basu}}},
  \bibinfo{author}{\bibfnamefont{J.}~\bibnamefont{{Chluba}}},
  \bibinfo{author}{\bibfnamefont{J.}~\bibnamefont{{Sayers}}},
  \bibinfo{author}{\bibfnamefont{R.}~\bibnamefont{{Adam}}},
  \bibinfo{author}{\bibfnamefont{E.}~\bibnamefont{{Churazov}}},
  \bibinfo{author}{\bibfnamefont{A.}~\bibnamefont{{Crites}}},
  \bibinfo{author}{\bibfnamefont{L.}~\bibnamefont{{Di Mascolo}}},
  \bibinfo{author}{\bibfnamefont{D.}~\bibnamefont{{Eckert}}},
  \bibnamefont{et~al.}, \bibinfo{journal}{\ssr} \textbf{\bibinfo{volume}{215}},
  \bibinfo{eid}{17} (\bibinfo{year}{2019}), \eprint{1811.02310}.

\bibitem[{\citenamefont{{Kompaneets}}(1957)}]{1957JETP...4.730K}
\bibinfo{author}{\bibfnamefont{A.~S.} \bibnamefont{{Kompaneets}}},
  \bibinfo{journal}{Soviet Journal of Experimental and Theoretical Physics}
  \textbf{\bibinfo{volume}{4}}, \bibinfo{pages}{730} (\bibinfo{year}{1957}).

\bibitem[{\citenamefont{{Chluba}
  et~al.}(2012{\natexlab{a}})\citenamefont{{Chluba}, {Khatri}, and
  {Sunyaev}}}]{2012MNRAS.425.1129C}
\bibinfo{author}{\bibfnamefont{J.}~\bibnamefont{{Chluba}}},
  \bibinfo{author}{\bibfnamefont{R.}~\bibnamefont{{Khatri}}}, \bibnamefont{and}
  \bibinfo{author}{\bibfnamefont{R.~A.} \bibnamefont{{Sunyaev}}},
  \bibinfo{journal}{\mnras} \textbf{\bibinfo{volume}{425}},
  \bibinfo{pages}{1129} (\bibinfo{year}{2012}{\natexlab{a}}),
  \eprint{1202.0057}.

\bibitem[{\citenamefont{{Chluba} et~al.}(2014)\citenamefont{{Chluba}, {Dai},
  and {Kamionkowski}}}]{2014MNRAS.437...67C}
\bibinfo{author}{\bibfnamefont{J.}~\bibnamefont{{Chluba}}},
  \bibinfo{author}{\bibfnamefont{L.}~\bibnamefont{{Dai}}}, \bibnamefont{and}
  \bibinfo{author}{\bibfnamefont{M.}~\bibnamefont{{Kamionkowski}}},
  \bibinfo{journal}{\mnras} \textbf{\bibinfo{volume}{437}}, \bibinfo{pages}{67}
  (\bibinfo{year}{2014}), \eprint{1308.5969}.

\bibitem[{\citenamefont{{Chluba} and {Dai}}(2014)}]{2014MNRAS.438.1324C}
\bibinfo{author}{\bibfnamefont{J.}~\bibnamefont{{Chluba}}} \bibnamefont{and}
  \bibinfo{author}{\bibfnamefont{L.}~\bibnamefont{{Dai}}},
  \bibinfo{journal}{\mnras} \textbf{\bibinfo{volume}{438}},
  \bibinfo{pages}{1324} (\bibinfo{year}{2014}), \eprint{1309.3274}.

\bibitem[{\citenamefont{{Challinor} and {Lasenby}}(1998)}]{1998ApJ...499....1C}
\bibinfo{author}{\bibfnamefont{A.}~\bibnamefont{{Challinor}}} \bibnamefont{and}
  \bibinfo{author}{\bibfnamefont{A.}~\bibnamefont{{Lasenby}}},
  \bibinfo{journal}{\apj} \textbf{\bibinfo{volume}{499}}, \bibinfo{pages}{1}
  (\bibinfo{year}{1998}), \eprint{astro-ph/9711161}.

\bibitem[{\citenamefont{{Itoh} et~al.}(1998)\citenamefont{{Itoh}, {Kohyama},
  and {Nozawa}}}]{1998ApJ...502....7I}
\bibinfo{author}{\bibfnamefont{N.}~\bibnamefont{{Itoh}}},
  \bibinfo{author}{\bibfnamefont{Y.}~\bibnamefont{{Kohyama}}},
  \bibnamefont{and} \bibinfo{author}{\bibfnamefont{S.}~\bibnamefont{{Nozawa}}},
  \bibinfo{journal}{\apj} \textbf{\bibinfo{volume}{502}}, \bibinfo{pages}{7}
  (\bibinfo{year}{1998}), \eprint{astro-ph/9712289}.

\bibitem[{\citenamefont{Stebbins}(1997)}]{Stebbins:1997qr}
\bibinfo{author}{\bibfnamefont{A.}~\bibnamefont{Stebbins}},
  \bibinfo{journal}{Submitted to: Astrophys. J. Lett.}  (\bibinfo{year}{1997}),
  \eprint{astro-ph/9709065}.

\bibitem[{\citenamefont{Shimon and Rephaeli}(2002)}]{Rephaeli:2002zs}
\bibinfo{author}{\bibfnamefont{M.}~\bibnamefont{Shimon}} \bibnamefont{and}
  \bibinfo{author}{\bibfnamefont{Y.}~\bibnamefont{Rephaeli}},
  \bibinfo{journal}{Astrophys. J.} \textbf{\bibinfo{volume}{575}},
  \bibinfo{pages}{12} (\bibinfo{year}{2002}), \eprint{astro-ph/0204355}.

\bibitem[{\citenamefont{{Itoh} et~al.}(2000{\natexlab{a}})\citenamefont{{Itoh},
  {Kawana}, {Nozawa}, and {Kohyama}}}]{2000astro.ph..5390I}
\bibinfo{author}{\bibfnamefont{N.}~\bibnamefont{{Itoh}}},
  \bibinfo{author}{\bibfnamefont{Y.}~\bibnamefont{{Kawana}}},
  \bibinfo{author}{\bibfnamefont{S.}~\bibnamefont{{Nozawa}}}, \bibnamefont{and}
  \bibinfo{author}{\bibfnamefont{Y.}~\bibnamefont{{Kohyama}}},
  \bibinfo{journal}{ArXiv Astrophysics e-prints}
  (\bibinfo{year}{2000}{\natexlab{a}}), \eprint{astro-ph/0005390}.

\bibitem[{\citenamefont{{Hu} et~al.}(1994)\citenamefont{{Hu}, {Scott}, and
  {Silk}}}]{1994PhRvD..49..648H}
\bibinfo{author}{\bibfnamefont{W.}~\bibnamefont{{Hu}}},
  \bibinfo{author}{\bibfnamefont{D.}~\bibnamefont{{Scott}}}, \bibnamefont{and}
  \bibinfo{author}{\bibfnamefont{J.}~\bibnamefont{{Silk}}},
  \bibinfo{journal}{\prd} \textbf{\bibinfo{volume}{49}}, \bibinfo{pages}{648}
  (\bibinfo{year}{1994}), \eprint{astro-ph/9305038}.

\bibitem[{\citenamefont{{Nozawa} et~al.}(1998)\citenamefont{{Nozawa}, {Itoh},
  and {Kohyama}}}]{1998ApJ...508...17N}
\bibinfo{author}{\bibfnamefont{S.}~\bibnamefont{{Nozawa}}},
  \bibinfo{author}{\bibfnamefont{N.}~\bibnamefont{{Itoh}}}, \bibnamefont{and}
  \bibinfo{author}{\bibfnamefont{Y.}~\bibnamefont{{Kohyama}}},
  \bibinfo{journal}{\apj} \textbf{\bibinfo{volume}{508}}, \bibinfo{pages}{17}
  (\bibinfo{year}{1998}), \eprint{astro-ph/9804051}.

\bibitem[{\citenamefont{{Sazonov} and {Sunyaev}}(1998)}]{1998ApJ...508....1S}
\bibinfo{author}{\bibfnamefont{S.~Y.} \bibnamefont{{Sazonov}}}
  \bibnamefont{and} \bibinfo{author}{\bibfnamefont{R.~A.}
  \bibnamefont{{Sunyaev}}}, \bibinfo{journal}{\apj}
  \textbf{\bibinfo{volume}{508}}, \bibinfo{pages}{1} (\bibinfo{year}{1998}).

\bibitem[{\citenamefont{{Challinor} and {Lasenby}}(1999)}]{1999ApJ...510..930C}
\bibinfo{author}{\bibfnamefont{A.}~\bibnamefont{{Challinor}}} \bibnamefont{and}
  \bibinfo{author}{\bibfnamefont{A.}~\bibnamefont{{Lasenby}}},
  \bibinfo{journal}{\apj} \textbf{\bibinfo{volume}{510}}, \bibinfo{pages}{930}
  (\bibinfo{year}{1999}), \eprint{astro-ph/9805329}.

\bibitem[{\citenamefont{{Nozawa} and {Kohyama}}(2009)}]{2009PhRvD..79h3005N}
\bibinfo{author}{\bibfnamefont{S.}~\bibnamefont{{Nozawa}}} \bibnamefont{and}
  \bibinfo{author}{\bibfnamefont{Y.}~\bibnamefont{{Kohyama}}},
  \bibinfo{journal}{\prd} \textbf{\bibinfo{volume}{79}}, \bibinfo{eid}{083005}
  (\bibinfo{year}{2009}), \eprint{0902.2595}.

\bibitem[{\citenamefont{{Nozawa} and {Kohyama}}(2013)}]{2013MNRAS.434..710N}
\bibinfo{author}{\bibfnamefont{S.}~\bibnamefont{{Nozawa}}} \bibnamefont{and}
  \bibinfo{author}{\bibfnamefont{Y.}~\bibnamefont{{Kohyama}}},
  \bibinfo{journal}{\mnras} \textbf{\bibinfo{volume}{434}},
  \bibinfo{pages}{710} (\bibinfo{year}{2013}), \eprint{1303.2286}.

\bibitem[{\citenamefont{{Nozawa} and {Kohyama}}(2014)}]{2014MNRAS.441.3018N}
\bibinfo{author}{\bibfnamefont{S.}~\bibnamefont{{Nozawa}}} \bibnamefont{and}
  \bibinfo{author}{\bibfnamefont{Y.}~\bibnamefont{{Kohyama}}},
  \bibinfo{journal}{\mnras} \textbf{\bibinfo{volume}{441}},
  \bibinfo{pages}{3018} (\bibinfo{year}{2014}), \eprint{1402.1541}.

\bibitem[{\citenamefont{{Chluba}
  et~al.}(2012{\natexlab{b}})\citenamefont{{Chluba}, {Nagai}, {Sazonov}, and
  {Nelson}}}]{2012MNRAS.426..510C}
\bibinfo{author}{\bibfnamefont{J.}~\bibnamefont{{Chluba}}},
  \bibinfo{author}{\bibfnamefont{D.}~\bibnamefont{{Nagai}}},
  \bibinfo{author}{\bibfnamefont{S.}~\bibnamefont{{Sazonov}}},
  \bibnamefont{and} \bibinfo{author}{\bibfnamefont{K.}~\bibnamefont{{Nelson}}},
  \bibinfo{journal}{\mnras} \textbf{\bibinfo{volume}{426}},
  \bibinfo{pages}{510} (\bibinfo{year}{2012}{\natexlab{b}}),
  \eprint{1205.5778}.

\bibitem[{\citenamefont{{Chluba} et~al.}(2005)\citenamefont{{Chluba},
  {H{\"u}tsi}, and {Sunyaev}}}]{2005A&A...434..811C}
\bibinfo{author}{\bibfnamefont{J.}~\bibnamefont{{Chluba}}},
  \bibinfo{author}{\bibfnamefont{G.}~\bibnamefont{{H{\"u}tsi}}},
  \bibnamefont{and} \bibinfo{author}{\bibfnamefont{R.~A.}
  \bibnamefont{{Sunyaev}}}, \bibinfo{journal}{\aap}
  \textbf{\bibinfo{volume}{434}}, \bibinfo{pages}{811} (\bibinfo{year}{2005}),
  \eprint{astro-ph/0409058}.

\bibitem[{\citenamefont{{Lavaux} et~al.}(2004)\citenamefont{{Lavaux}, {Diego},
  {Mathis}, and {Silk}}}]{2004MNRAS.347..729L}
\bibinfo{author}{\bibfnamefont{G.}~\bibnamefont{{Lavaux}}},
  \bibinfo{author}{\bibfnamefont{J.~M.} \bibnamefont{{Diego}}},
  \bibinfo{author}{\bibfnamefont{H.}~\bibnamefont{{Mathis}}}, \bibnamefont{and}
  \bibinfo{author}{\bibfnamefont{J.}~\bibnamefont{{Silk}}},
  \bibinfo{journal}{\mnras} \textbf{\bibinfo{volume}{347}},
  \bibinfo{pages}{729} (\bibinfo{year}{2004}), \eprint{astro-ph/0307293}.

\bibitem[{\citenamefont{{Sazonov} and {Sunyaev}}(1999)}]{1999MNRAS.310..765S}
\bibinfo{author}{\bibfnamefont{S.~Y.} \bibnamefont{{Sazonov}}}
  \bibnamefont{and} \bibinfo{author}{\bibfnamefont{R.~A.}
  \bibnamefont{{Sunyaev}}}, \bibinfo{journal}{\mnras}
  \textbf{\bibinfo{volume}{310}}, \bibinfo{pages}{765} (\bibinfo{year}{1999}),
  \eprint{astro-ph/9903287}.

\bibitem[{\citenamefont{{Kamionkowski} and {Loeb}}(1997)}]{1997PhRvD..56.4511K}
\bibinfo{author}{\bibfnamefont{M.}~\bibnamefont{{Kamionkowski}}}
  \bibnamefont{and} \bibinfo{author}{\bibfnamefont{A.}~\bibnamefont{{Loeb}}},
  \bibinfo{journal}{\prd} \textbf{\bibinfo{volume}{56}}, \bibinfo{pages}{4511}
  (\bibinfo{year}{1997}), \eprint{astro-ph/9703118}.

\bibitem[{\citenamefont{{Challinor} et~al.}(2000)\citenamefont{{Challinor},
  {Ford}, and {Lasenby}}}]{2000MNRAS.312..159C}
\bibinfo{author}{\bibfnamefont{A.~D.} \bibnamefont{{Challinor}}},
  \bibinfo{author}{\bibfnamefont{M.~T.} \bibnamefont{{Ford}}},
  \bibnamefont{and} \bibinfo{author}{\bibfnamefont{A.~N.}
  \bibnamefont{{Lasenby}}}, \bibinfo{journal}{\mnras}
  \textbf{\bibinfo{volume}{312}}, \bibinfo{pages}{159} (\bibinfo{year}{2000}),
  \eprint{astro-ph/9905227}.

\bibitem[{\citenamefont{{Shehzad Emritte} et~al.}(2016)\citenamefont{{Shehzad
  Emritte}, {Colafrancesco}, and {Marchegiani}}}]{2016JCAP...07..031S}
\bibinfo{author}{\bibfnamefont{M.}~\bibnamefont{{Shehzad Emritte}}},
  \bibinfo{author}{\bibfnamefont{S.}~\bibnamefont{{Colafrancesco}}},
  \bibnamefont{and}
  \bibinfo{author}{\bibfnamefont{P.}~\bibnamefont{{Marchegiani}}},
  \bibinfo{journal}{Journal of Cosmology and Astro-Particle Physics}
  \textbf{\bibinfo{volume}{2016}}, \bibinfo{eid}{031} (\bibinfo{year}{2016}).

\bibitem[{\citenamefont{{Itoh} et~al.}(2000{\natexlab{b}})\citenamefont{{Itoh},
  {Nozawa}, and {Kohyama}}}]{2000ApJ...533..588I}
\bibinfo{author}{\bibfnamefont{N.}~\bibnamefont{{Itoh}}},
  \bibinfo{author}{\bibfnamefont{S.}~\bibnamefont{{Nozawa}}}, \bibnamefont{and}
  \bibinfo{author}{\bibfnamefont{Y.}~\bibnamefont{{Kohyama}}},
  \bibinfo{journal}{\apj} \textbf{\bibinfo{volume}{533}}, \bibinfo{pages}{588}
  (\bibinfo{year}{2000}{\natexlab{b}}), \eprint{astro-ph/9812376}.

\bibitem[{\citenamefont{{Yasini} and {Pierpaoli}}(2016)}]{2016PhRvD..94b3513Y}
\bibinfo{author}{\bibfnamefont{S.}~\bibnamefont{{Yasini}}} \bibnamefont{and}
  \bibinfo{author}{\bibfnamefont{E.}~\bibnamefont{{Pierpaoli}}},
  \bibinfo{journal}{\prd} \textbf{\bibinfo{volume}{94}}, \bibinfo{eid}{023513}
  (\bibinfo{year}{2016}).

\bibitem[{\citenamefont{{Edigaryev} et~al.}(2018)\citenamefont{{Edigaryev},
  {Novikov}, and {Pilipenko}}}]{2018PhRvD..98l3513E}
\bibinfo{author}{\bibfnamefont{I.~G.} \bibnamefont{{Edigaryev}}},
  \bibinfo{author}{\bibfnamefont{D.~I.} \bibnamefont{{Novikov}}},
  \bibnamefont{and} \bibinfo{author}{\bibfnamefont{S.~V.}
  \bibnamefont{{Pilipenko}}}, \bibinfo{journal}{\prd}
  \textbf{\bibinfo{volume}{98}}, \bibinfo{eid}{123513} (\bibinfo{year}{2018}),
  \eprint{1812.01330}.

\bibitem[{\citenamefont{{Babuel-Peyrissac} and
  {Rouvillois}}(1969)}]{1969JPhys..30..301B}
\bibinfo{author}{\bibfnamefont{J.~P.} \bibnamefont{{Babuel-Peyrissac}}}
  \bibnamefont{and}
  \bibinfo{author}{\bibfnamefont{G.}~\bibnamefont{{Rouvillois}}},
  \bibinfo{journal}{Journal de Physique} \textbf{\bibinfo{volume}{30}},
  \bibinfo{pages}{301} (\bibinfo{year}{1969}).

\bibitem[{\citenamefont{{Pomraning}}(1974)}]{1974ApJ...191..183P}
\bibinfo{author}{\bibfnamefont{G.~C.} \bibnamefont{{Pomraning}}},
  \bibinfo{journal}{\apj} \textbf{\bibinfo{volume}{191}}, \bibinfo{pages}{183}
  (\bibinfo{year}{1974}).

\bibitem[{\citenamefont{{Stark}}(1981)}]{1981MNRAS.195..115S}
\bibinfo{author}{\bibfnamefont{R.~F.} \bibnamefont{{Stark}}},
  \bibinfo{journal}{\mnras} \textbf{\bibinfo{volume}{195}},
  \bibinfo{pages}{115} (\bibinfo{year}{1981}).

\bibitem[{\citenamefont{{Efstathiou}}(2003)}]{2003MNRAS.346L..26E}
\bibinfo{author}{\bibfnamefont{G.}~\bibnamefont{{Efstathiou}}},
  \bibinfo{journal}{\mnras} \textbf{\bibinfo{volume}{346}},
  \bibinfo{pages}{L26} (\bibinfo{year}{2003}), \eprint{astro-ph/0306431}.

\bibitem[{\citenamefont{{Tegmark} et~al.}(2003)\citenamefont{{Tegmark}, {de
  Oliveira-Costa}, and {Hamilton}}}]{2003PhRvD..68l3523T}
\bibinfo{author}{\bibfnamefont{M.}~\bibnamefont{{Tegmark}}},
  \bibinfo{author}{\bibfnamefont{A.}~\bibnamefont{{de Oliveira-Costa}}},
  \bibnamefont{and} \bibinfo{author}{\bibfnamefont{A.~J.}
  \bibnamefont{{Hamilton}}}, \bibinfo{journal}{\prd}
  \textbf{\bibinfo{volume}{68}}, \bibinfo{eid}{123523} (\bibinfo{year}{2003}),
  \eprint{astro-ph/0302496}.

\bibitem[{\citenamefont{{Schwarz} et~al.}(2004)\citenamefont{{Schwarz},
  {Starkman}, {Huterer}, and {Copi}}}]{2004PhRvL..93v1301S}
\bibinfo{author}{\bibfnamefont{D.~J.} \bibnamefont{{Schwarz}}},
  \bibinfo{author}{\bibfnamefont{G.~D.} \bibnamefont{{Starkman}}},
  \bibinfo{author}{\bibfnamefont{D.}~\bibnamefont{{Huterer}}},
  \bibnamefont{and} \bibinfo{author}{\bibfnamefont{C.~J.}
  \bibnamefont{{Copi}}}, \bibinfo{journal}{Physical Review Letters}
  \textbf{\bibinfo{volume}{93}}, \bibinfo{eid}{221301} (\bibinfo{year}{2004}),
  \eprint{astro-ph/0403353}.

\bibitem[{\citenamefont{{Planck Collaboration}
  et~al.}(2014)\citenamefont{{Planck Collaboration}, {Ade}, {Aghanim},
  {Armitage-Caplan}, {Arnaud}, {Ashdown}, {Atrio-Barandela}, {Aumont},
  {Baccigalupi}, {Banday} et~al.}}]{2014A&A...571A..15P}
\bibinfo{author}{\bibnamefont{{Planck Collaboration}}},
  \bibinfo{author}{\bibfnamefont{P.~A.~R.} \bibnamefont{{Ade}}},
  \bibinfo{author}{\bibfnamefont{N.}~\bibnamefont{{Aghanim}}},
  \bibinfo{author}{\bibfnamefont{C.}~\bibnamefont{{Armitage-Caplan}}},
  \bibinfo{author}{\bibfnamefont{M.}~\bibnamefont{{Arnaud}}},
  \bibinfo{author}{\bibfnamefont{M.}~\bibnamefont{{Ashdown}}},
  \bibinfo{author}{\bibfnamefont{F.}~\bibnamefont{{Atrio-Barandela}}},
  \bibinfo{author}{\bibfnamefont{J.}~\bibnamefont{{Aumont}}},
  \bibinfo{author}{\bibfnamefont{C.}~\bibnamefont{{Baccigalupi}}},
  \bibinfo{author}{\bibfnamefont{A.~J.} \bibnamefont{{Banday}}},
  \bibnamefont{et~al.}, \bibinfo{journal}{\aap} \textbf{\bibinfo{volume}{571}},
  \bibinfo{eid}{A15} (\bibinfo{year}{2014}), \eprint{1303.5075}.

\bibitem[{\citenamefont{{Copi} et~al.}(2004)\citenamefont{{Copi}, {Huterer},
  and {Starkman}}}]{2004PhRvD..70d3515C}
\bibinfo{author}{\bibfnamefont{C.~J.} \bibnamefont{{Copi}}},
  \bibinfo{author}{\bibfnamefont{D.}~\bibnamefont{{Huterer}}},
  \bibnamefont{and} \bibinfo{author}{\bibfnamefont{G.~D.}
  \bibnamefont{{Starkman}}}, \bibinfo{journal}{\prd}
  \textbf{\bibinfo{volume}{70}}, \bibinfo{eid}{043515} (\bibinfo{year}{2004}),
  \eprint{astro-ph/0310511}.

\bibitem[{\citenamefont{{Copi} et~al.}(2006)\citenamefont{{Copi}, {Huterer},
  {Schwarz}, and {Starkman}}}]{2006MNRAS.367...79C}
\bibinfo{author}{\bibfnamefont{C.~J.} \bibnamefont{{Copi}}},
  \bibinfo{author}{\bibfnamefont{D.}~\bibnamefont{{Huterer}}},
  \bibinfo{author}{\bibfnamefont{D.~J.} \bibnamefont{{Schwarz}}},
  \bibnamefont{and} \bibinfo{author}{\bibfnamefont{G.~D.}
  \bibnamefont{{Starkman}}}, \bibinfo{journal}{\mnras}
  \textbf{\bibinfo{volume}{367}}, \bibinfo{pages}{79} (\bibinfo{year}{2006}),
  \eprint{astro-ph/0508047}.

\bibitem[{\citenamefont{{Naselsky} and
  {Verkhodanov}}(2008)}]{2008IJMPD..17..179N}
\bibinfo{author}{\bibfnamefont{P.~D.} \bibnamefont{{Naselsky}}}
  \bibnamefont{and} \bibinfo{author}{\bibfnamefont{O.~V.}
  \bibnamefont{{Verkhodanov}}}, \bibinfo{journal}{International Journal of
  Modern Physics D} \textbf{\bibinfo{volume}{17}}, \bibinfo{pages}{179}
  (\bibinfo{year}{2008}), \eprint{astro-ph/0609409}.

\bibitem[{\citenamefont{{Sachs} and {Wolfe}}(1967)}]{1967ApJ...147...73S}
\bibinfo{author}{\bibfnamefont{R.~K.} \bibnamefont{{Sachs}}} \bibnamefont{and}
  \bibinfo{author}{\bibfnamefont{A.~M.} \bibnamefont{{Wolfe}}},
  \bibinfo{journal}{\apj} \textbf{\bibinfo{volume}{147}}, \bibinfo{pages}{73}
  (\bibinfo{year}{1967}).

\bibitem[{\citenamefont{{Wild} et~al.}(2009)\citenamefont{{Wild}, {Kardashev},
  {Likhachev}, {Babakin}, {Arkhipov}, {Vinogradov}, {Andreyanov}, {Fedorchuk},
  {Myshonkova}, {Alexsandrov} et~al.}}]{2009ExA....23..221W}
\bibinfo{author}{\bibfnamefont{W.}~\bibnamefont{{Wild}}},
  \bibinfo{author}{\bibfnamefont{N.~S.} \bibnamefont{{Kardashev}}},
  \bibinfo{author}{\bibfnamefont{S.~F.} \bibnamefont{{Likhachev}}},
  \bibinfo{author}{\bibfnamefont{N.~G.} \bibnamefont{{Babakin}}},
  \bibinfo{author}{\bibfnamefont{V.~Y.} \bibnamefont{{Arkhipov}}},
  \bibinfo{author}{\bibfnamefont{I.~S.} \bibnamefont{{Vinogradov}}},
  \bibinfo{author}{\bibfnamefont{V.~V.} \bibnamefont{{Andreyanov}}},
  \bibinfo{author}{\bibfnamefont{S.~D.} \bibnamefont{{Fedorchuk}}},
  \bibinfo{author}{\bibfnamefont{N.~V.} \bibnamefont{{Myshonkova}}},
  \bibinfo{author}{\bibfnamefont{Y.~A.} \bibnamefont{{Alexsandrov}}},
  \bibnamefont{et~al.}, \bibinfo{journal}{Experimental Astronomy}
  \textbf{\bibinfo{volume}{23}}, \bibinfo{pages}{221} (\bibinfo{year}{2009}).

\bibitem[{\citenamefont{{Kardashev} et~al.}(2014)\citenamefont{{Kardashev},
  {Novikov}, {Lukash}, {Pilipenko}, {Mikheeva}, {Bisikalo}, {Wiebe},
  {Doroshkevich}, {Zasov}, {Zinchenko} et~al.}}]{2014PhyU...57.1199K}
\bibinfo{author}{\bibfnamefont{N.~S.} \bibnamefont{{Kardashev}}},
  \bibinfo{author}{\bibfnamefont{I.~D.} \bibnamefont{{Novikov}}},
  \bibinfo{author}{\bibfnamefont{V.~N.} \bibnamefont{{Lukash}}},
  \bibinfo{author}{\bibfnamefont{S.~V.} \bibnamefont{{Pilipenko}}},
  \bibinfo{author}{\bibfnamefont{E.~V.} \bibnamefont{{Mikheeva}}},
  \bibinfo{author}{\bibfnamefont{D.~V.} \bibnamefont{{Bisikalo}}},
  \bibinfo{author}{\bibfnamefont{D.~S.} \bibnamefont{{Wiebe}}},
  \bibinfo{author}{\bibfnamefont{A.~G.} \bibnamefont{{Doroshkevich}}},
  \bibinfo{author}{\bibfnamefont{A.~V.} \bibnamefont{{Zasov}}},
  \bibinfo{author}{\bibfnamefont{I.~I.} \bibnamefont{{Zinchenko}}},
  \bibnamefont{et~al.}, \bibinfo{journal}{Physics Uspekhi}
  \textbf{\bibinfo{volume}{57}}, \bibinfo{eid}{1199-1228}
  (\bibinfo{year}{2014}), \eprint{1502.06071}.

\bibitem[{\citenamefont{{Smirnov} et~al.}(2012)\citenamefont{{Smirnov},
  {Baryshev}, {Pilipenko}, {Myshonkova}, {Bulanov}, {Arkhipov}, {Vinogradov},
  {Likhachev}, and {Kardashev}}}]{2012SPIE.8442E..4CS}
\bibinfo{author}{\bibfnamefont{A.~V.} \bibnamefont{{Smirnov}}},
  \bibinfo{author}{\bibfnamefont{A.~M.} \bibnamefont{{Baryshev}}},
  \bibinfo{author}{\bibfnamefont{S.~V.} \bibnamefont{{Pilipenko}}},
  \bibinfo{author}{\bibfnamefont{N.~V.} \bibnamefont{{Myshonkova}}},
  \bibinfo{author}{\bibfnamefont{V.~B.} \bibnamefont{{Bulanov}}},
  \bibinfo{author}{\bibfnamefont{M.~Y.} \bibnamefont{{Arkhipov}}},
  \bibinfo{author}{\bibfnamefont{I.~S.} \bibnamefont{{Vinogradov}}},
  \bibinfo{author}{\bibfnamefont{S.~F.} \bibnamefont{{Likhachev}}},
  \bibnamefont{and} \bibinfo{author}{\bibfnamefont{N.~S.}
  \bibnamefont{{Kardashev}}}, in \emph{\bibinfo{booktitle}{Space Telescopes and
  Instrumentation 2012: Optical, Infrared, and Millimeter Wave}}
  (\bibinfo{year}{2012}), vol. \bibinfo{volume}{8442} of
  \emph{\bibinfo{series}{Proc. SPIE}}, p. \bibinfo{pages}{84424C}.

\bibitem[{\citenamefont{{Zeldovich} and
  {Sunyaev}}(1969{\natexlab{b}})}]{1969Ap&SS...4..301}
\bibinfo{author}{\bibfnamefont{Y.~B.} \bibnamefont{{Zeldovich}}}
  \bibnamefont{and} \bibinfo{author}{\bibfnamefont{R.~A.}
  \bibnamefont{{Sunyaev}}}, \bibinfo{journal}{\apss}
  \textbf{\bibinfo{volume}{4}}, \bibinfo{pages}{301}
  (\bibinfo{year}{1969}{\natexlab{b}}).

\bibitem[{\citenamefont{{Illarionov} and
  {Siuniaev}}(1974)}]{1974AZh....51.1162I}
\bibinfo{author}{\bibfnamefont{A.~F.} \bibnamefont{{Illarionov}}}
  \bibnamefont{and} \bibinfo{author}{\bibfnamefont{R.~A.}
  \bibnamefont{{Siuniaev}}}, \bibinfo{journal}{\azh}
  \textbf{\bibinfo{volume}{51}}, \bibinfo{pages}{1162} (\bibinfo{year}{1974}).

\bibitem[{\citenamefont{{Burigana} et~al.}(1991)\citenamefont{{Burigana},
  {Danese}, and {de Zotti}}}]{1991A&A...246...49B}
\bibinfo{author}{\bibfnamefont{C.}~\bibnamefont{{Burigana}}},
  \bibinfo{author}{\bibfnamefont{L.}~\bibnamefont{{Danese}}}, \bibnamefont{and}
  \bibinfo{author}{\bibfnamefont{G.}~\bibnamefont{{de Zotti}}},
  \bibinfo{journal}{\aap} \textbf{\bibinfo{volume}{246}}, \bibinfo{pages}{49}
  (\bibinfo{year}{1991}).

\bibitem[{\citenamefont{{Hu} and {Silk}}(1993)}]{1993PhRvD..48..485H}
\bibinfo{author}{\bibfnamefont{W.}~\bibnamefont{{Hu}}} \bibnamefont{and}
  \bibinfo{author}{\bibfnamefont{J.}~\bibnamefont{{Silk}}},
  \bibinfo{journal}{\prd} \textbf{\bibinfo{volume}{48}}, \bibinfo{pages}{485}
  (\bibinfo{year}{1993}).

\bibitem[{\citenamefont{{Chluba} and {Sunyaev}}(2012)}]{2012MNRAS.419.1294C}
\bibinfo{author}{\bibfnamefont{J.}~\bibnamefont{{Chluba}}} \bibnamefont{and}
  \bibinfo{author}{\bibfnamefont{R.~A.} \bibnamefont{{Sunyaev}}},
  \bibinfo{journal}{\mnras} \textbf{\bibinfo{volume}{419}},
  \bibinfo{pages}{1294} (\bibinfo{year}{2012}), \eprint{1109.6552}.

\bibitem[{\citenamefont{{Chluba}}(2015)}]{2015MNRAS.454.4182C}
\bibinfo{author}{\bibfnamefont{J.}~\bibnamefont{{Chluba}}},
  \bibinfo{journal}{\mnras} \textbf{\bibinfo{volume}{454}},
  \bibinfo{pages}{4182} (\bibinfo{year}{2015}), \eprint{1506.06582}.

\bibitem[{\citenamefont{{Chluba}
  et~al.}(2019{\natexlab{a}})\citenamefont{{Chluba}, {Kogut}, {Patil},
  {Abitbol}, {Aghanim}, {Ali-Ha{\i}{\ensuremath{\ddot{}}}moud}, {Amin},
  {Aumont}, {Bartolo}, {Basu} et~al.}}]{2019BAAS...51c.184C}
\bibinfo{author}{\bibfnamefont{J.}~\bibnamefont{{Chluba}}},
  \bibinfo{author}{\bibfnamefont{A.}~\bibnamefont{{Kogut}}},
  \bibinfo{author}{\bibfnamefont{S.~P.} \bibnamefont{{Patil}}},
  \bibinfo{author}{\bibfnamefont{M.~H.} \bibnamefont{{Abitbol}}},
  \bibinfo{author}{\bibfnamefont{N.}~\bibnamefont{{Aghanim}}},
  \bibinfo{author}{\bibfnamefont{Y.}~\bibnamefont{{Ali-Ha{\i}{\ensuremath{\ddot{}}}moud}}},
  \bibinfo{author}{\bibfnamefont{M.~A.} \bibnamefont{{Amin}}},
  \bibinfo{author}{\bibfnamefont{J.}~\bibnamefont{{Aumont}}},
  \bibinfo{author}{\bibfnamefont{N.}~\bibnamefont{{Bartolo}}},
  \bibinfo{author}{\bibfnamefont{K.}~\bibnamefont{{Basu}}},
  \bibnamefont{et~al.}, \bibinfo{journal}{\baas} \textbf{\bibinfo{volume}{51}},
  \bibinfo{eid}{184} (\bibinfo{year}{2019}{\natexlab{a}}), \eprint{1903.04218}.

\bibitem[{\citenamefont{{Chluba}
  et~al.}(2019{\natexlab{b}})\citenamefont{{Chluba}, {Abitbol}, {Aghanim},
  {Ali-Haimoud}, {Alvarez}, {Basu}, {Bolliet}, {Burigana}, {de Bernardis},
  {Delabrouille} et~al.}}]{2019arXiv190901593C}
\bibinfo{author}{\bibfnamefont{J.}~\bibnamefont{{Chluba}}},
  \bibinfo{author}{\bibfnamefont{M.~H.} \bibnamefont{{Abitbol}}},
  \bibinfo{author}{\bibfnamefont{N.}~\bibnamefont{{Aghanim}}},
  \bibinfo{author}{\bibfnamefont{Y.}~\bibnamefont{{Ali-Haimoud}}},
  \bibinfo{author}{\bibfnamefont{M.}~\bibnamefont{{Alvarez}}},
  \bibinfo{author}{\bibfnamefont{K.}~\bibnamefont{{Basu}}},
  \bibinfo{author}{\bibfnamefont{B.}~\bibnamefont{{Bolliet}}},
  \bibinfo{author}{\bibfnamefont{C.}~\bibnamefont{{Burigana}}},
  \bibinfo{author}{\bibfnamefont{P.}~\bibnamefont{{de Bernardis}}},
  \bibinfo{author}{\bibfnamefont{J.}~\bibnamefont{{Delabrouille}}},
  \bibnamefont{et~al.}, \bibinfo{journal}{arXiv e-prints}
  \bibinfo{eid}{arXiv:1909.01593} (\bibinfo{year}{2019}{\natexlab{b}}),
  \eprint{1909.01593}.

\bibitem[{\citenamefont{{Desjacques} et~al.}(2015)\citenamefont{{Desjacques},
  {Chluba}, {Silk}, {de Bernardis}, and {Dor{\'e}}}}]{2015MNRAS.451.4460D}
\bibinfo{author}{\bibfnamefont{V.}~\bibnamefont{{Desjacques}}},
  \bibinfo{author}{\bibfnamefont{J.}~\bibnamefont{{Chluba}}},
  \bibinfo{author}{\bibfnamefont{J.}~\bibnamefont{{Silk}}},
  \bibinfo{author}{\bibfnamefont{F.}~\bibnamefont{{de Bernardis}}},
  \bibnamefont{and}
  \bibinfo{author}{\bibfnamefont{O.}~\bibnamefont{{Dor{\'e}}}},
  \bibinfo{journal}{\mnras} \textbf{\bibinfo{volume}{451}},
  \bibinfo{pages}{4460} (\bibinfo{year}{2015}), \eprint{1503.05589}.

\bibitem[{\citenamefont{{Chluba}}(2014)}]{2014arXiv1405.6938C}
\bibinfo{author}{\bibfnamefont{J.}~\bibnamefont{{Chluba}}},
  \bibinfo{journal}{ArXiv e-prints} \bibinfo{eid}{arXiv:1405.6938}
  (\bibinfo{year}{2014}), \eprint{1405.6938}.

\bibitem[{\citenamefont{{Fixsen}}(2009)}]{2009ApJ...707..916F}
\bibinfo{author}{\bibfnamefont{D.~J.} \bibnamefont{{Fixsen}}},
  \bibinfo{journal}{\apj} \textbf{\bibinfo{volume}{707}}, \bibinfo{pages}{916}
  (\bibinfo{year}{2009}), \eprint{0911.1955}.

\bibitem[{\citenamefont{{Crittenden} and {Turok}}(1996)}]{1996PhRvL..76..575C}
\bibinfo{author}{\bibfnamefont{R.~G.} \bibnamefont{{Crittenden}}}
  \bibnamefont{and} \bibinfo{author}{\bibfnamefont{N.}~\bibnamefont{{Turok}}},
  \bibinfo{journal}{Physical Review Letters} \textbf{\bibinfo{volume}{76}},
  \bibinfo{pages}{575} (\bibinfo{year}{1996}), \eprint{astro-ph/9510072}.

\bibitem[{\citenamefont{{Planck Collaboration}
  et~al.}(2018)\citenamefont{{Planck Collaboration}, {Akrami}, {Ashdown},
  {Aumont}, {Baccigalupi}, {Ballardini}, {Band ay}, {Barreiro}, {Bartolo},
  {Basak} et~al.}}]{2018arXiv180706208P}
\bibinfo{author}{\bibnamefont{{Planck Collaboration}}},
  \bibinfo{author}{\bibfnamefont{Y.}~\bibnamefont{{Akrami}}},
  \bibinfo{author}{\bibfnamefont{M.}~\bibnamefont{{Ashdown}}},
  \bibinfo{author}{\bibfnamefont{J.}~\bibnamefont{{Aumont}}},
  \bibinfo{author}{\bibfnamefont{C.}~\bibnamefont{{Baccigalupi}}},
  \bibinfo{author}{\bibfnamefont{M.}~\bibnamefont{{Ballardini}}},
  \bibinfo{author}{\bibfnamefont{A.~J.} \bibnamefont{{Band ay}}},
  \bibinfo{author}{\bibfnamefont{R.~B.} \bibnamefont{{Barreiro}}},
  \bibinfo{author}{\bibfnamefont{N.}~\bibnamefont{{Bartolo}}},
  \bibinfo{author}{\bibfnamefont{S.}~\bibnamefont{{Basak}}},
  \bibnamefont{et~al.}, \bibinfo{journal}{arXiv e-prints}
  \bibinfo{eid}{arXiv:1807.06208} (\bibinfo{year}{2018}), \eprint{1807.06208}.

\bibitem[{\citenamefont{{Planck Collaboration}
  et~al.}(2016)\citenamefont{{Planck Collaboration}, {Ade}, {Aghanim},
  {Arnaud}, {Ashdown}, {Aumont}, {Baccigalupi}, {Banday}, {Barreiro}, {Barrena}
  et~al.}}]{2016A&A...594A..27P}
\bibinfo{author}{\bibnamefont{{Planck Collaboration}}},
  \bibinfo{author}{\bibfnamefont{P.~A.~R.} \bibnamefont{{Ade}}},
  \bibinfo{author}{\bibfnamefont{N.}~\bibnamefont{{Aghanim}}},
  \bibinfo{author}{\bibfnamefont{M.}~\bibnamefont{{Arnaud}}},
  \bibinfo{author}{\bibfnamefont{M.}~\bibnamefont{{Ashdown}}},
  \bibinfo{author}{\bibfnamefont{J.}~\bibnamefont{{Aumont}}},
  \bibinfo{author}{\bibfnamefont{C.}~\bibnamefont{{Baccigalupi}}},
  \bibinfo{author}{\bibfnamefont{A.~J.} \bibnamefont{{Banday}}},
  \bibinfo{author}{\bibfnamefont{R.~B.} \bibnamefont{{Barreiro}}},
  \bibinfo{author}{\bibfnamefont{R.}~\bibnamefont{{Barrena}}},
  \bibnamefont{et~al.}, \bibinfo{journal}{\aap} \textbf{\bibinfo{volume}{594}},
  \bibinfo{eid}{A27} (\bibinfo{year}{2016}), \eprint{1502.01598}.

\end{thebibliography}



\end{document}